\documentclass[conference]{IEEEtran}

\usepackage{cite}
\usepackage [
  n,
  advantage,
  operators,
  sets,
  adversary,
  landau,
  probability,
  notions,
  logic,
  ff,
  mm,
  primitives,
  events,
  complexity,
  asymptotics,
  keys
] {cryptocode}
\usepackage{amsmath,amssymb,amsfonts}
\usepackage{algorithmic}
\usepackage{chronology}
\usepackage{listings}
\usepackage{graphicx}
\usepackage{textcomp}
\usepackage{xcolor}
\usepackage{makecell}
\usepackage{hyperref}
\usepackage{caption}
\usepackage{subcaption}
\usepackage{courier}
\usepackage{float}
\usepackage{amsmath}
\usepackage{url}

\title{TrustZero - open, verifiable and scalable zero-trust
}
\author{\IEEEauthorblockN{Adrian-Tudor Dumitrescu}
\IEEEauthorblockA{ 
\textit{Delft University of Technology}\\
Delft, The Netherlands \\
A.T.Dumitrescu@student.tudelft.nl}
\and
\IEEEauthorblockN{Johan Pouwelse }
\IEEEauthorblockA{
\textit{Delft University of Technology}\\
Delft, The Netherlands \\
J.A.Pouwelse@tudelft.nl}}

\begin{document}

\maketitle

\thispagestyle{plain}

\begin{abstract}
We present a passport-level trust token for Europe. In an era of escalating cyber threats fueled by global competition in economic, military, and technological domains, traditional security models are proving inadequate. The rise of advanced attacks exploiting zero-day vulnerabilities, supply chain infiltration, and system interdependencies underscores the need for a paradigm shift in cybersecurity. Zero Trust Architecture (ZTA) emerges as a transformative framework that replaces implicit trust with continuous verification of identity and granular access control. This thesis introduces TrustZero, a scalable layer of zero-trust security built around a universal "trust token" - a non-revocable self-sovereign identity with cryptographic signatures to enable robust, mathematically grounded trust attestations. By integrating ZTA principles with cryptography, TrustZero establishes a secure web-of-trust framework adaptable to legacy systems and inter-organisational communication.
\end{abstract}

\section{Introduction}

In an era marked by intense global competition across economic, military, and technological spheres, the digital landscape has become a critical battleground. Nations and organizations worldwide are investing heavily in cyber capabilities to gain a strategic advantage, leading to a rise in cyber threats that target both government and private sectors \cite{petratos2014cybersecurity}. 
A major concern in this landscape is the presence of “zero-day” vulnerabilities—previously unknown security flaws in software or hardware that lack any available defenses \cite{fidler2014anarchy}. These vulnerabilities are highly valuable, often traded on a global market and exploited by state actors and criminal groups to infiltrate systems, steal sensitive data, and disrupt operations.
One such example is the Israeli NSO Group that used spyware Pegasus for remote zero-click surveillance of smartphones for goals "aligned with the geopolitical interests"\cite{kaster2023privatized}.
Zero Trust Architecture has emerged as a response to this escalating arms race in cybersecurity. 

Traditional security models that rely on perimeter-based defenses, such as Virtual Private Networks (VPNs) or firewalls, are proving inadequate against advanced, multi-vector cyber threats. This shift is underscored by the European Systemic Risk Board's findings\cite{eubook}, which highlight a persistently heightened cyber threat landscape in Europe with sabotage of underwater telecommunications cables and disruption to systems in major financial institutions.
In this sector, cyber risks have evolved in tandem with these threats. Attackers have become adept at exploiting complex system interdependencies to maximize damage, compelling financial institutions to elevate their security stance. Programs like the TIBER-NL\cite{tiberNL} (Threat Intelligence-based Ethical Red-teaming) in the Netherlands are part of a proactive approach where regulated firms undergo simulated, controlled cyberattacks based on real-world threat scenarios. At the European level, the European Central Bank runs a similar initiative to ensure systemic stability by exposing potential vulnerabilities through rigorous, scenario-based testing \cite{tiberEU}. Even the United Nations Security Council acknowledges the digital world has become a favorable field for espionage and cyberterrorism that creates "mistrust and paranoia between nations"\cite{santonisecurity}. 

The proliferation of hybrid warfare has steadily penetrated societal activities, with even elections becoming a primary target for digital disruption. These attacks have been studied since the 2014 Scottish elections where "fictional accounts of conspiracy theories" \cite{birch2017fraud} spread misinformation and fear among citizens. This pattern of interferences peaked in 2024 with the Romanian presidential elections where cyberattacks sought to exploit vulnerabilities in the election IT system and influence people through fake accounts. This triggered the EU to search for a solution regarding bot activity and fraud by giving social media company TikTok a "retention order that concerns national elections"\cite{europaCommissionOnline}. Secret documents have been declassified to the public\cite{politicoRomaniasPresidential} and presented 25,000 accounts as part of a network on TikTok that became very active in the two weeks before the elections. From those, around 800 had existed since 2016, the year TikTok was released, but with almost no activity until November of this year. The Romanian Secret Services also observed that each TikTok account was associated with a unique IP address, indicating a deliberate strategy to obscure the true scale of the attack.
The European Commission followed the press release with a formal proceeding against the company to "assess and mitigate systemic risks" with the commission's president stating that "foreign actors interfered in the Romanian presidential elections by using TikTok"\cite{europaCommissionOpens}.

Zero Trust Architecture (ZTA) has emerged as a modern security framework grounded in continuous verification of identity and contextual access requests. ZTA, proposed by Kindervag in 2010 \cite{kindervag2010build}, follows the core principle that trust must never be assumed and that every user or device must be verified regardless. This new vision of the Internet prevents lateral movement intrusion, a type of attack encountered by big tech companies, based on the assumption that threats are omnipresent and no traffic can be trusted, including from internal networks \cite{9585170}.

The criticality of a Zero Trust Architecture is further highlighted by high-profile breaches, such as Google’s “Operation Aurora” incident in 2010. In this attack, Chinese state-sponsored actors exploited a zero-day vulnerability to infiltrate not only the original company but also Adobe and over 30 other major corporations\cite{OCONNOR2013125}. The Operation Aurora case emphasized the risk posed by supply chain infiltration, where attackers compromise secondary suppliers of defense contractors to gain access to sensitive information. By exploiting the interconnected nature of global supply chains, adversaries can effectively bypass direct defenses, underscoring the need for a zero-trust approach that assumes no entity is trustworthy by default, regardless of its location within or outside the organization’s network. Another such example of attack is Titan Rain, started in 2003 and targeting defense contractor computer networks in the US\cite{rogin2007cyber}. The attack source was identified as Guangdong, China, where perpetrators constantly changed IP addresses to make tracking their movements harder.
 
Current researches focus on moving from single-factor to multi-factor and continuous authentication, improving security while minimizing resource use \cite{he2022survey}, but most projects stop at the concept phase. This trend started with solutions developed by giant tech companies like Google's BeyondCorp. Motivated by the security breaches presented above, the company started a workflow transition to a protected zero-trust network, migration that proved to be harder than expected\cite{gonccalves2023beyondcorp}. 

This thesis presents \textbf{TrustZero}, a layer of zero-trust security designed to be applied at large scale. As a proof-of-principle we created a universal "trust token" that consists of a non-revocable self-sovereign identity with a list of trust attestations. The architecture relies on the "trust token" flow between clients and servers during communication. We present a trust model/algorithm that is rigorously based on mathematical axioms with verifiable cryptographic signatures. By combining the zero-trust principles of continuous verification and cryptography we create a strong identity and web-of-trust framework that can serve as an upgrade for legacy communication infrastructure that can be applied between organisations. TrustZero is founded on the innovative principle of enabling trust to be both portable and verifiable across a global scale. In a world where the "hybrid" war is escalating with denial of service and election interferences, we consider TrustZero would substantially make such attacks more difficult, costly and less effective.

\begin{figure}
\begin{subfigure}{.25\textwidth}
  \centering
  \includegraphics[width=.9\linewidth]{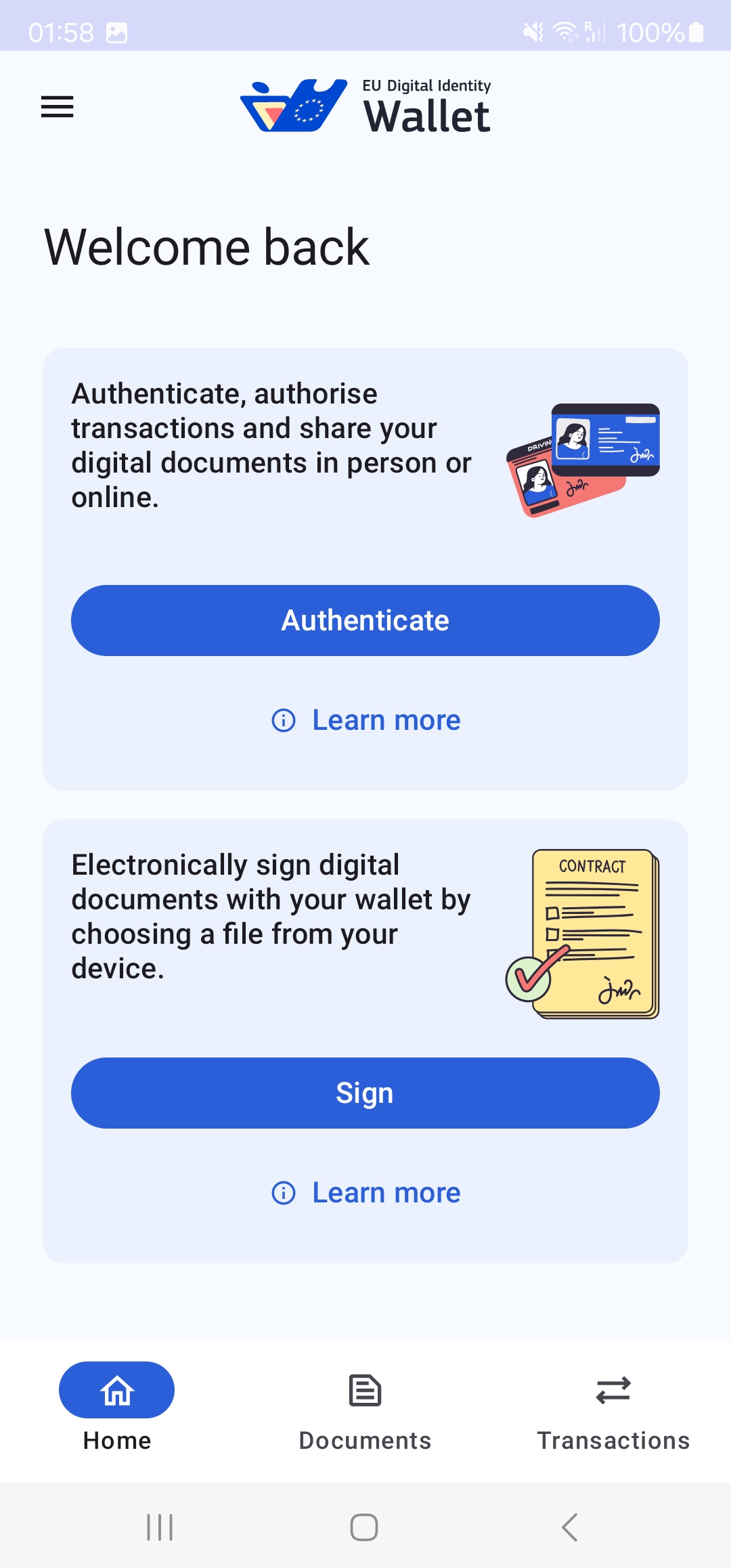}
  \caption{EUDI home screen}
  \label{fig:eudi_home}
\end{subfigure}%
\begin{subfigure}{.25\textwidth}
  \centering
  \includegraphics[width=.9\linewidth]{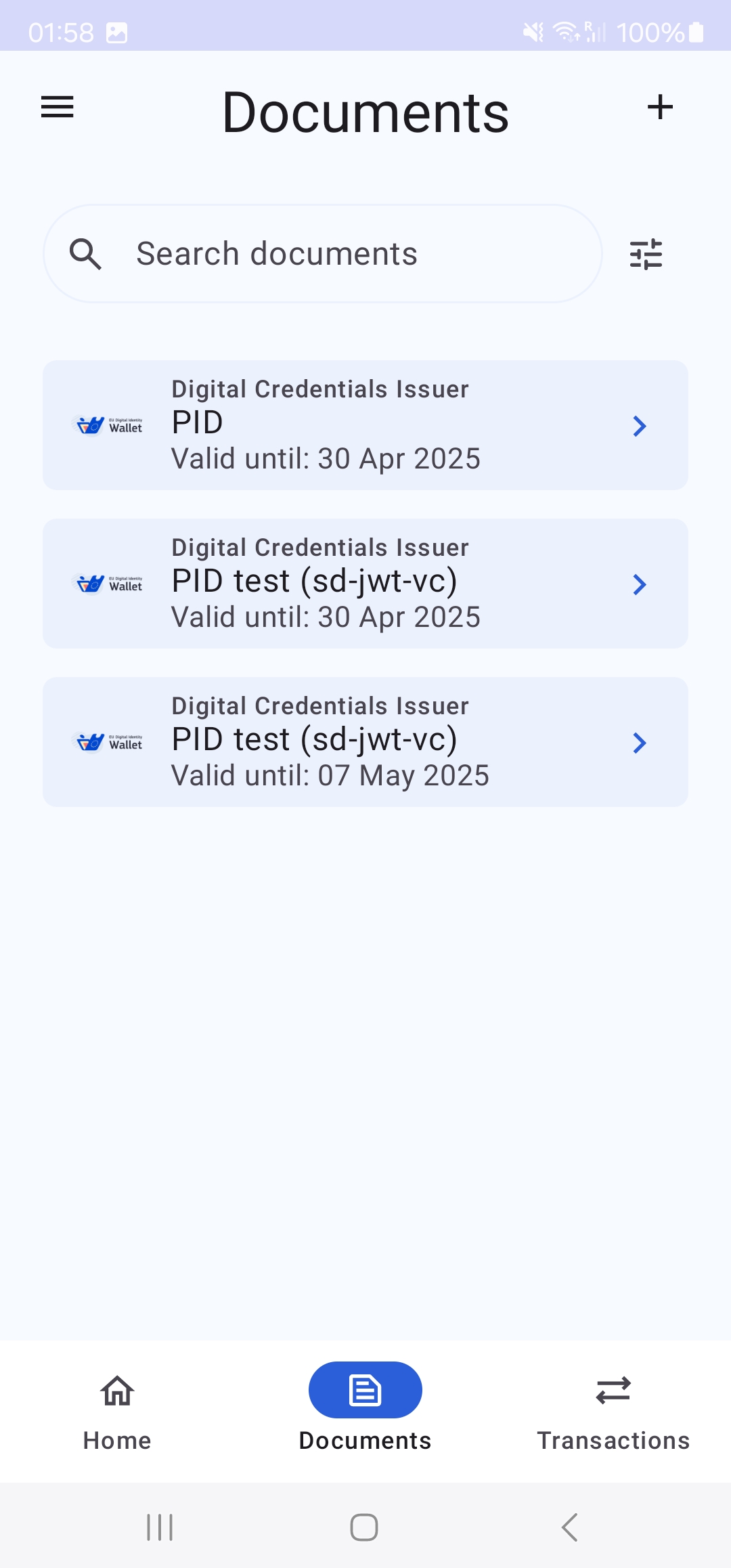}
  \caption{EUDI Digital identities}
  \label{fig:eudi_id}
\end{subfigure}
\caption{EUDI wallet \cite{eudi}}
\label{fig:eudi_app}
\end{figure}

\begin{figure*}[h]
  \includegraphics[width=.8\linewidth]{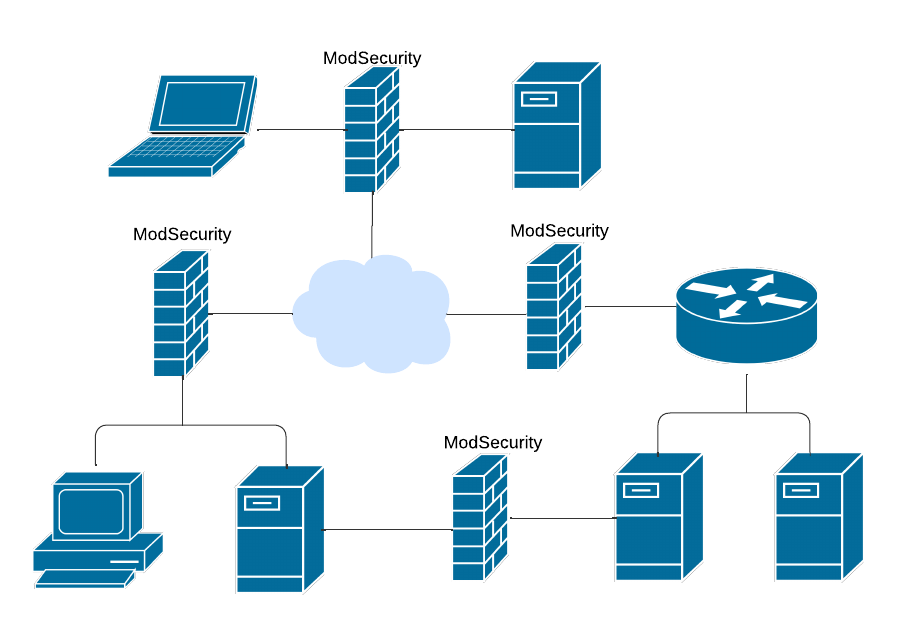}
  \centering
  \caption{TrustZero architecture overview}
  \label{fig:architecture}
\end{figure*}

\section{The fundaments of Zero Trust Architecture}
Zero Trust Architecture fundamentally redefines cybersecurity by shifting from implicit trust to \textbf{continuous verification}. Every user and device must be authenticated and authorized before accessing network resources, reducing the risk of lateral threats. This model enhances data security and detects anomalies, making it a vital framework in today’s cloud-driven, remote-access world. A key aspect of Zero Trust is its direct linkage to the first principles of digital identity, where the emphasis is placed on verifying the authenticity and integrity of digital identities in real-time. This means that identity is treated as the new perimeter, and digital identity verification becomes a cornerstone of security. ZTA helps ensure that only authenticated and authorized users or devices, with minimal privileges, can access critical resources, key requirements to provide passport-grade identity. TrustZero was developed to adhere to these requirements and to complement the European Digital Identity Wallet, presented in Figure \ref{fig:eudi_app}, with a communicable, tamper-proof and verifiable token of trust.

Building a zero-trust architecture starts from five basic assumptions that can solve the security problems encountered in large-scale supply-chain networks:
\begin{enumerate}
    \item The network is constantly exposed to a hostile environment.

    \item Threats, both internal and external nodes, persist throughout the network's operation.

    \item A network's location alone cannot determine its trustworthiness.

    \item Every device, user, and network traffic must be continuously authenticated and authorized.

    \item Security policies need to be adaptable and dynamically recalculated based on a wide range of data inputs.
\end{enumerate}

According to the National Institute of Standards and Technology (NIST), ZTA relies on three core logical components to enforce security policies. The Policy Enforcement Point (PEP) is the first of these, acting as an intermediary between the user and the server; it enables, monitors, and eventually terminates the connection between the subject and the resource, creating a boundary often referred to as the trust-zone. Closely collaborating with the PEP is the Policy Administrator (PA), which is responsible for granting or denying access based on the PEP’s assessments. Finally, the Policy Engine (PE) functions as the “brain” of the system, making access decisions by applying a trust algorithm to external inputs, in alignment with the organization’s security policies\cite{stafford2020zero}.

In a digital identity-focused system, the policy engine (PE) plays a pivotal role in making access decisions according to business strategies. The trust computation is considered the cornerstone for access control in a zero-trust architecture and there is no universal algorithm agreed. Different approaches have been tried, from score-based/weight-based evaluations to fuzzy logic or graph theory model\cite{9773102} but most of them prove to be too complex and hard to deploy on large-scale systems. Migrating to a zero-trust architecture is a complex task, as highlighted in \cite{teerakanok2021migrating}. Most trust computations rely on external information and when a provider encounters security or technical issues, finding a quick replacement can be challenging and costly. Requesting resources in a ZTA environment requires calculating a trust level based on credentials and information provided by the requester, then comparing it to the predetermined trust level required for accessing specific resources. Establishing an appropriate trust level for each resource is a complex task. The enterprise must find a balance in the trust levels—too high can make resources difficult to access, potentially hindering workflows, while too low may result in inadequate security. The trust algorithm represents the most common bottleneck regarding resource usage where complex dynamic scoring is based on multiple factors: login, network and operational behavior, user and device identities, user behavior, terminal security status and risks assessment (\cite{chen2020security}, \cite{yao2020dynamic}, \cite{vanickis2018access}).

One of the main concerns regarding ZTA is that, in fact, zero-trust is an impossible property to achieve for a system. As highlighted in a recent work\cite{gligor2022zero}, "Zero Trust" is fundamentally unachievable for certain security properties in "black box" devices—systems whose internal operations can not be fully inspected. Specifically, malware in such devices cannot be definitively ruled out, as verifications of security properties can only offer probabilistic assurances rather than certainties. Thus, complete trust elimination is impossible, and trust establishment becomes the practical alternative. Moreover, ZTA is considered impractical, demanding high assurance for all security properties of network devices, meaning their correctness must be proven rigorously. Instead, defenders focus on mitigating the cost of breaches rather than preventing them entirely, as low assurance and breaches are inevitable in real-world systems. This problem was reinforced by work in 2010\cite{6062066}, expressing that, in security, trust can only be relocated not established. In relation to the impossibility of zero-trust, Lampson \cite{lampson2009privacy} stated that each part of prevention architecture is complex as a whole and there are always new threats, making security a fractal. Policy in such a system "it has to be simple", "it has to minimize hassle for the user, at least most of the time" and "it has to be true (given some assumptions)", assumptions hard to achieve in ZTA.

\section{Problem description}

The field of zero trust security faces significant challenges related to transparency, openness, standardization, and reproducibility. Despite its growing adoption and the crucial role it plays in protecting sensitive infrastructures, the field lacks an open-source reference architecture or implementation that demonstrates exemplary zero trust security practices. This limitation extends to scientific research, where reproducibility is a cornerstone but remains underdeveloped in the context of zero trust systems. The absence of detailed descriptions of attack methodologies, defensive software architectures, security tooling, and policies contributes to an opaque landscape. This is true even in academia, where cybersecurity research and practice are often secluded. For instance, at institutions like Delft University, while other computer science faculties are accessible, cybersecurity departments maintain restricted access, highlighting an ingrained culture of secrecy. This is reflected in the methodologies in zero-trust where projects like OpenZiti\cite{openzitiWhatOpenZiti} are using cryptographic principles such as low-cost hardware security modules (HSMs)\cite{openzitiHardwareSecurity} that leverage the laws of physics.

Moreover, the current body of scientific literature on zero trust is sparse in providing comprehensive, reproducible case studies or implementations. The seminal example of Google's response to the successful cyberattack Aurora illustrates this gap. Following the breach, a zero-trust model has been employed to isolate the legal, financial, engineering, and administrative networks. However, their documentation and public disclosures are described only at a high level(\cite{ward2014beyondcorp}, \cite{escobedo2017beyondcorp}, \cite{peck2017migrating}), lacking the granular details necessary for academic analysis and reproducibility.

Addressing this gap is critical not just for advancing zero trust methodologies and fostering an environment of shared knowledge but also to withstand the demands of modern cybersecurity. As stated before, the global competition is altering the digital landscape with attacks on critical infrastructure of finance and healthcare in vulnerable moments like the COVID pandemic\cite{chigada2021cyberattacks} and started to interfere even in election systems. Moreover, the attack traffic is carried by many autonomous systems (AS) along the way, indirectly assisting illegitimate traffic, based on the assumption that all packets are trustworthy. There are multiple studies to detect and prevent malicious traffic through the network (\cite{thapngam2011discriminating}, \cite{feghhi2016web}, \cite{meghdouri2018analysis}), but are based on specific patterns and scenarios(data-driven) and most require extensive computational time/power.

In contrast, Zero Trust Architecture emphasizes \textit{continuous verification of every user and device, regardless} of their location within the network involving regularly reassessing the trustworthiness of all participants. Even though this architecture was introduced in 2010, to this day, no real-world open-source reference applications have been made \cite{he2022survey}. Thus, the overarching challenge lies in developing a comprehensive, open-source zero trust framework with full transparency that is easy to reproduced in sensitive infrastructure.

\section{Our TrustZero design}

To address the limitations of existing zero trust implementations, we designed TrustZero, a framework that leverages complexity as a strength rather than a vulnerability. This architecture is designed for large and complex supply chains bringing new trust between organisations. Traditional approaches often regard complexity as a source of fragility. This exposes systems to faults and weaknesses as demonstrated by critical incidents such as the exposed Log4j vulnerability\cite{hiesgen2022race}. Moreover, major disruptions can occur like the CrowdStrike 2024 global outage\cite{ogundipe2024shaky} determined by an intricate interaction with Windows systems. These incidents forced major organizations to focus "more on proactive defense strategies, away from the traditional perimeter-based protection to continuous monitoring of the internal systems"\cite{naseer2024crowdstrike}. In various high-stakes industries, including defense contracting, aerospace, and semiconductor manufacturing, complexity presents formidable challenges that can undermine system reliability. The semiconductor industry leader ASML, for example, relies on an intricate network of over 5,000 suppliers\cite{asmlResponsibleSupply}—a testament to the depth and diversity inherent in modern supply chains. This complexity can also be observed in the automotive industry where companies like Audi recognize their responsibility to maintain a network of 14,000 of direct suppliers in over 60 countries\cite{audiResponsibilitySupply}.

TrustZero capitalizes on this intricate connectivity, redefining zero trust principles to operate beyond the confines of a single entity, organization, or government. Unlike traditional zero trust models that focus solely on securing an isolated system or entity, TrustZero is built on the novel principle of making trust portable and verifiable at a global scale. This approach facilitates trust exchange between various autonomous entities, enabling a unified, interoperable trust network. Collective intelligence is a unique and impactful way organizations work together, made possible by our architecture. For instance, approaches such as Byzantine-robust learning with compression\cite{rammal2024communication} could be shielded by TrustZero from misinformation, deception, spam, and fraud. Exploring this direction is left as future work.
More exactly, our key contributions are as follows:
\begin{enumerate}
    \item We propose an architecture that is based on transparency and reproducibility of simple cryptographic functions to create trust between nodes. The protocol increases the security of communication to servers by adding client trust score evaluation.
    \item We provide an open-source minimum-viable product
    of the architecture to demonstrate the correctness and
    functionality.
\end{enumerate}

At the core of TrustZero is the concept of shared and transparent authentication histories, which support collaborative verification processes. By making successful interactions publicly traceable, systems develop greater resilience through emergent patterns and verifiable trust. This resilience is especially pronounced when critical components of the authentication network use high-assurance, passport-grade verification mechanisms, such as those compliant with eIDAS (Electronic Identification, Authentication and Trust Services) standards. This ensures that even machine-to-machine (M2M) communications are protected by strong, interoperable trust assurance. This links, as noted in a 2014 ASML co-authored publication \cite{frigieri2015m2m}, are traditionally implemented using protocols like MQTT that were not designed with inherent security and are known in being vulnerable\cite{dinculeanua2019vulnerabilities}.

As a proof-of-concept, we developed a universal trust token, featuring a self-sovereign identity with non-revocable trust attestations. We further enhanced a web application firewall, ModSecurity, with trust scoring, real-time threat signaling, and collaboration and placed it as a reverse proxy and gateway for each server present in the network to form a global web-of-trust.
Lastly, we present a global trust model rooted in mathematical(cryptographic) principles, combining the zero-trust elements to create a robust identity and trust framework. The high-overview architecture of TrustZero is represented in Figure \ref{fig:architecture}.

\subsection{Protocol}
The TrustZero Protocol exemplifies a decentralized approach to secure trust verification in multi-party systems, addressing critical challenges in integrity and authenticity. In this protocol, a user generates a public/private keys ($(\pk_{u}, \sk_{u}) \gets \kgen$) and sequentially interacts with multiple servers, each generating its own pair ($(\pk_{sn}, \sk_{sn}) \gets \kgen$). The user initiates communication by sending messages and their public key to the servers, where each server signs the user’s public key using its private key, producing signatures ($sign_N \gets \sign(\sk_{sN}, \pk_{u})$). Each server processes 2 sequential steps ensuring both correctness and accountability:
\begin{enumerate}
    \item verifies all the previously generated signatures by other servers
    \item issues its own signature if the previous step passed(or is renewing it)
\end{enumerate}
The resulting "trust token", a concatenation of all signatures, serves as verifiable proof of trust, validated against the servers' public keys. By distributing the signature generation and verification processes, the protocol eliminates the reliance on a central authority, enhancing security and trustworthiness. The protocol’s integration of cryptographic keys, digital signatures, and decentralized verification provides a robust framework for securing trust in systems vulnerable to adversarial threats, making it suitable for applications in distributed supply-chain networks and secure communications.

\subsection{Trust token}
In our zero trust architecture, the trust token is the key element that enables a non-revocable, self-sovereign identity with a list of trust attestations. In this identity model, users maintain full ownership and control of their digital identity without third-party oversight or the risk of revocation. This identity is accompanied by a collection of trust attestations—verified endorsements or credentials—that validate the user’s identity and reputation.

In a self-sovereign, zero-trust architecture, users maintain complete control of their trust tokens without relying on external cloud storage. The core design choice lies between transparent trust communication and an explicit trust protocol. Trust is communicated passively by embedding a session request header and token signatures can be verified and added as a layer of security without altering existing server protocols.

The list of trust attestations present in a trust token is a series of signatures that the user received from past (adequate) interactions with other nodes. These signatures can be verified by any server with simple cryptographic functions against the public key of the user with the public key of the issuing server. In exchange for a "good" request, the server will issue a new trust token for the user with a new signature (or update the present one if they already interacted).
The trust token sent in each request will be structured in the form:
\begin{equation}
token = sign_{server_1}||sign_{server_2}||...||sign_{server_N}
\end{equation}

The EUDI app from Figure \ref{fig:eudi_app} uses a traditional JSON web token to exchange information and verify identities for access. This approach proves to be rigid regarding token revocation\cite{janoky2018analysis} with a bad actor being able to use it even after detection. The trust token used by TrustZero can be revoked by any server after any abnormal access, lowering the user's reputation. Moreover, a server can change its key pair to invalidate all its signatures and indirectly signal the user to all other servers that will interact with it.

\subsection{Trust algorithm}

The trust algorithm implemented in TrustZero is score-based, evaluating the trustworthiness of entities and assigning a numerical score that reflects their reliability. The value is computed by verifying the number of valid signatures a user has on its public key. The crucial objective is to make it as expensive as possible to set up, maintain, and/or exit fake identities and give priority to trusted parties. To achieve this, the architecture exploits a resource that most attackers do not have: time. With TrustZero scoring system, trust is gained over time and interaction with unique servers, reducing the surprise of an attack (like denial of service for example).

The trust computation is constructed on basic cryptographic functions that require low resources and are easy to understand, from administration entities to simple users. This algorithm is executed before a request even gets to a server not interfering with its normal behavior. In the end, the TrustZero algorithm for a user ($\pk_{U}$) that has signatures from n servers($S$) can be summarized as:
\begin{multline}
trustscore = Vf(sign_{S_1}, \pk_{S_1}, \pk_{U}) + \\ + Vf(sign_{S_2}, \pk_{S_2}, \pk_{U}) + ... \\ ... + Vf(sign_{S_N}, \pk_{S_N}, \pk_{U})
\end{multline}
where $Vf()$ is a cryptographic function that returns 0 or 1 based on the validation of a signature using the public key of the originator server against the public key of the user.

\subsection{ModSecurity}
ModSecurity\cite{ristic2010modsecurity} is an open-source web application firewall (WAF) used to monitor, log, and filter HTTP traffic to prevent attacks on web applications. It acts as a security layer between users and web servers by inspecting requests and responses based on customizable rules, helping to detect and block threats such as SQL injection, cross-site scripting (XSS), and other vulnerabilities. ModSecurity can also be used for real-time monitoring, auditing, and compliance with security standards, making it a vital tool for enhancing web application security.

In TrustZero, ModSecurity is deployed in reverse proxy mode where it acts as an intermediary between clients and servers, inspecting all inbound and outbound traffic before it reaches the destination web server. This setup allows ModSecurity to enforce security rules, log traffic, and block attacks without modifying the web server itself, acting as a Policy enforcement point.

The capabilities of ModSecurity in detecting and mitigating different attacks have been extensively studied in state-of-the-art literature (\cite{8711673}, \cite{singh2018impact}, \cite{akbar2018sql}, \cite{lakhno2022experimental}). In our architecture, its most important characteristic is the non-disruptive nature of inspecting requests while enabling external scripting processing. The signatures present in a request are verified by ModSecurity which computes the score and can deny the communication (if signatures are wrong/compromised) or forward it to the servers. ModSecurity offers the perfect environment with its custom rules to adapt the security level(paranoia) for each request depending on its nature. While computing the trust score is studied in this thesis, its utilization in raising different security levels based on it is left for further research.

\section{Related work}

Although Zero Trust Architecture (ZTA) was first introduced over 10 years ago, no comprehensive, real-world implementation has emerged that fully addresses its potential. Most ZTA proposals remain in the design phase due to the complexity of their trust models and the need for substantial changes in infrastructure to accommodate them. Additionally, these solutions are often narrowly focused on specific business sectors, such as cloud computing, IoT etc, rather than providing a universal applicable framework applicable. This section will review these ZTA and security architecture proposals, highlighting the need for a more generalized, adaptable approach.

\subsection{First steps towards Zero-Trust Architecture}
TrustGuard\cite{srivatsa2005trustguard} model serves as an intermediary security layer that enforces strict access controls, monitors communication between entities, and validates interactions in real-time. Designed to reduce the risk of unauthorized access and lateral movement, TrustGuard bridges the gap between traditional network models and the zero-trust paradigm by establishing micro-boundaries of trust that are dynamically managed. It introduces a flow-level reputation-based defense mechanism and it was proposed as early as 2005 as a first step towards reputation and trust management networks. Unlike traditional methods that typically focus on IP addresses or individual packet characteristics, TrustGuard evaluates the reputation of entire network flows. Over time, it evolved in more specific uses cases such as allowing for more precise identification and mitigation of Distributed Denial of Service (DDoS) attack traffic\cite{liu2011trustguard} while reducing the incidence of false positives.

The architecture of TrustGuard encompasses several integral components. The flow collector is responsible for gathering detailed flow-level data, encompassing both traffic characteristics and behavioral patterns. The reputation manager analyzes this data to compute reputation scores for each flow, leveraging historical behavior alongside real-time observations. The decision engine then utilizes these reputation scores to make informed traffic filtering decisions, effectively distinguishing between legitimate and malicious flows. Additionally, a feedback loop continuously refines the reputation scores based on observed behaviors, enabling the system to adapt dynamically.

By incorporating machine learning techniques, TrustGuard enhances its ability to adaptively modify reputation scores in response to shifting traffic patterns and evolving attack characteristics. This combination of advanced analytics and real-time data processing positions TrustGuard as a robust solution for modern network security challenges.

\subsection{SDN and zero trust architecture}
A novel solution is presented in combining Software-defined networks(SDN) with zero-trust principles(\cite{guo2023intelligent}, \cite{zanasi2024flexible}), a security architecture designed to address the complex requirements of Industrial IoT systems, which include real-time operations, reliability, and decentralization. Traditional cybersecurity solutions struggle with the heterogeneity of IIoT devices. The proposed architectures leverages network micro-segmentation and integrates Software-Defined Networking (SDN) for policy enforcement, alongside a centralized security management layer for simplified control. A prototype demonstrates that this system ensures decentralized, resilient, and flexible security management while maintaining central oversight of security policies and network topology. One proposal \cite{zanasi2024flexible} uses Nebula, a software-defined overlay network solution, in an abstraction layer for policy enforcement. This tool relies on a custom Public Key Infrastructure (PKI) system since it uses certificates that are not X.509 compliant. 

Nebula introduces challenges in integrating with standard security frameworks and requires the development of a unique Certificate Authority (CA). Custom PKI solutions increase the complexity of managing certificate requests and key generation, already a demanding task, which may lead to security vulnerabilities. Additionally, isolating the Nebula network, while enhancing security, could introduce maintenance and scalability issues. Moreover, it is acknowledged that some devices might lack native support for Nebula and need to be integrated by introducing an additional device.

\subsection{BeyondCorp}
The "BeyondCorp"\cite{ward2014beyondcorp} model represents a paradigm shift in enterprise security, moving away from the traditional perimeter-centric approach. Developed by Google, it emphasizes user and device authentication regardless of location, allowing secure access to applications without a VPN. BeyondCorp relies on continuous verification through context-aware policies, integrating real-time monitoring and adaptive access controls to enhance security. The BeyondCorp model emphasizes secure device and user identification through a comprehensive management system. It maintains a Device Inventory Database to track managed devices, which are uniquely identified via device certificates stored in secure modules. User access is managed through a User and Group Database, integrated with HR processes, and authenticated via a Single Sign-On (SSO) system, which issues a session token for the access of a specific resource. Additionally, BeyondCorp establishes an unprivileged network that mimics an external network, enhancing security by minimizing trust in the internal network infrastructure.

As one of the few practical examples of ZTA, the model faced challenges during the later stages of Google's BeyondCorp migration, particularly regarding difficult use cases that did not fit the standard HTTPS-based workflow. Issues were signaled with specific applications that required IP-layer connectivity or could not easily integrate with the BeyondCorp access proxy\cite{gonccalves2023beyondcorp}.

\subsection{Zero trust in cloud computing}

Another discussed topic for ZTA is its use in cloud computing where services such as storage, processing power, databases, networking, software, and analytics are delivered over the internet. Thus, safeguarding such critical resources is key and the zero-trust design appears to satisfy the security requirements. For example, strategies with 9 principles of trust have been proposed but formulated just as a "conceptual model"\cite{9104214} for further research.

A novel concept presented for cloud computing is "survivable zero trust". Unlike existing models, the proposed architecture \cite{ferretti2021survivable} acknowledges that trusted components can be compromised. The novel survivable zero trust architecture ensures high security, robustness and can tolerate intrusions and recover from failures, making it suitable for cloud environments under specific conditions. The design is also based on a key pair and signature scheme that assists the communication against different attack scenarios. Even with a strong trust model, the paper recognizes that designing an effective protocol that ensures confidentiality while minimizing performance impacts and disruptions remains an open research challenge.

\subsection{Zero-trust and Blockchain}
Combining zero-trust and blockchain can enhance security in distributed systems addressing challenges such as identity management, secure data sharing, and ensuring compliance in decentralized environments. This movement materialized with projects like ZEBRA\cite{10646352}, a framework that focuses on securing Advanced Metering Infrastructure (AMI) using a Zero Trust Architecture combined with blockchain technology and Ring Oscillator Physical Unclonable Functions (ROPUFs). The design ensures robust device authentication and guarantees data integrity by leveraging the unique properties of ROPUFs, for generating unclonable keys, and blockchain for traceable and tamper-proof communication. This approach enhances the security of smart grid networks, providing resilience against cyber threats like unauthorized access, spoofing and data manipulation. Nevertheless, blockchain technology can introduce latency and require significant computational resources, which may be challenging for the resource-constrained devices used in AMI. Additionally, the reliance on ROPUFs for authentication, while secure, could be affected by environmental factors (such as temperature or voltage variations), impacting the reliability of the cryptographic keys generated. Managing these factors while maintaining system performance could pose challenges for real-world deployment.

Another solution, this time tailored for IoT is Amatista\cite{8473444}, a blockchain-based middleware designed for scalable management of IoT networks. The paper is the first to enumerate cryptography as an option in trust management but it incorporates it in the blockchain consensus algorithm. As IoT expands rapidly, the trustworthiness of millions of connected devices becomes a challenge. Amatista tackles this issue by employing a zero-trust approach, utilizing a novel hierarchical mining process to validate both the infrastructure and transactions at varying levels of trust. By leveraging blockchain features such as a distributed database, consensus mechanisms, smart contracts, and immutability, Amatista ensures reliable transactions without centralized validation nodes. The system is tested on Edison Arduino Boards, demonstrating how it can address trust concerns in IoT through decentralized validation mechanisms. While Amatista shows promise, potential issues include reliance on complex blockchain infrastructure, scalability challenges with numerous devices and, as stated by the authors, not yet tested "in a large scale loT deployment".

\section{Implementation and Experiments}

Our experiments focus on an in-depth exploration of real-world systems applying zero trust principles, combined with a comprehensive performance analysis of our novel TrustZero token. The key contributions of this work lie in demystifying the opaque security practices of major corporations and identifying the practical implications of deploying an open, verifiable zero trust system. Given the absence of a comprehensive zero trust solution that embodies end-to-end openness, open-source availability, and the potential for uncompromised self-hosting, our experiments are necessarily exploratory and integrative, evaluating various components to construct a robust, open framework.

\begin{figure}[h]
    \begin{lstlisting}[escapeinside={(*}{*)}]
    POST /resource HTTP/1.1
    Host: api.example.com
    Content-Type: application/json
    User-Key-Signatures: 
    (*$\pk_{user}$*):1:5d41402abc4b2a76b9719d91017c592
    Content-Length: 47
    {
      "username": "John Doe",
      "password": "johndoe",
    }
    \end{lstlisting}
    \caption{POST Request example}
  \label{fig:request}
\end{figure}

\begin{figure}[h]
  \includegraphics[width=.9\linewidth]{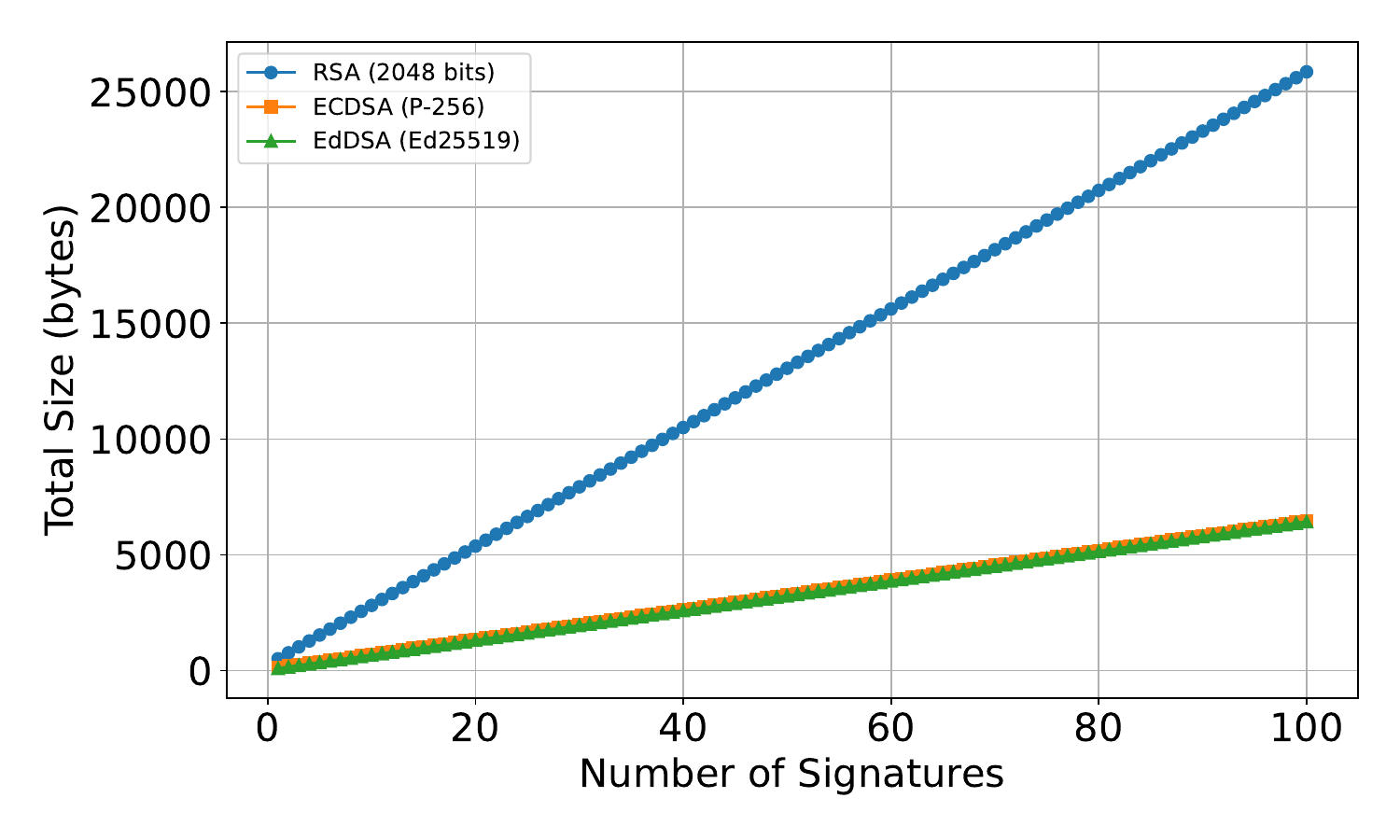}
  \centering
  \caption{Signatures size}
  \label{fig:sizes}
\end{figure}

The protocol outlined has been implemented in Python as a proof of concept. The open-source code is available on  \href{https://github.com/AdiDumi/TrustZero}{GitHub}\cite{reposit}. The Cryptography library is utilized for public/private key generation, as well as for signing and verification. A simple login server has also been created to accept (or deny) requests from users deployed alongside a ModSecurity instance acting as a reverse proxy for it. An example of POST request that was used in the experiments against the deployed server is presented in Figure \ref{fig:request}. The pair server-ModSecurity has been deployed with docker in multiple containers to simulate a distributed network of servers having their own ports and key pair. All traffic directed to the server is intercepted by ModSecurity who inspects the header of the request and searches for the trust token and public key. The rule related to this actions is defined in the following way:
\begin{lstlisting}
SecRule &REQUEST_HEADERS:User-Key-Signatures \
    "eq 0" \
    "id:10009, \
     phase:1, \
     t:none, \
     msg:'Missing User-Key-Signatures header', \
     deny"
\end{lstlisting}

In the case of a 'cold start', a user who has no signatures yet will only send his public key and will be treated accordingly with a score of 0.

Once the header is detected, ModSecurity inspects it to determine the number of valid signatures present and to asses the trust score of the user. The rule makes use of a Python script with cryptographic functions for verification and in case of an error signals ModSecurity to terminate the request:
\begin{lstlisting}
SecRule REQUEST_HEADERS:User-Key-Signatures \
    "@inspectFile /app/check_signatures.py" \
    "id:10010, \
     phase:1, \
     msg:'Error in signatures', \
     deny, \
     t:none"
\end{lstlisting}

If the request has all the correct signatures, it is passed to the server who can solve it accordingly. In the response header, the server attaches his signature to the token and passes it one more time through ModSecurity. After checking its presence, the response is forwarded to the user who can now store the signatures for further communication.
\begin{figure}[h]
  \includegraphics[width=.9\linewidth]{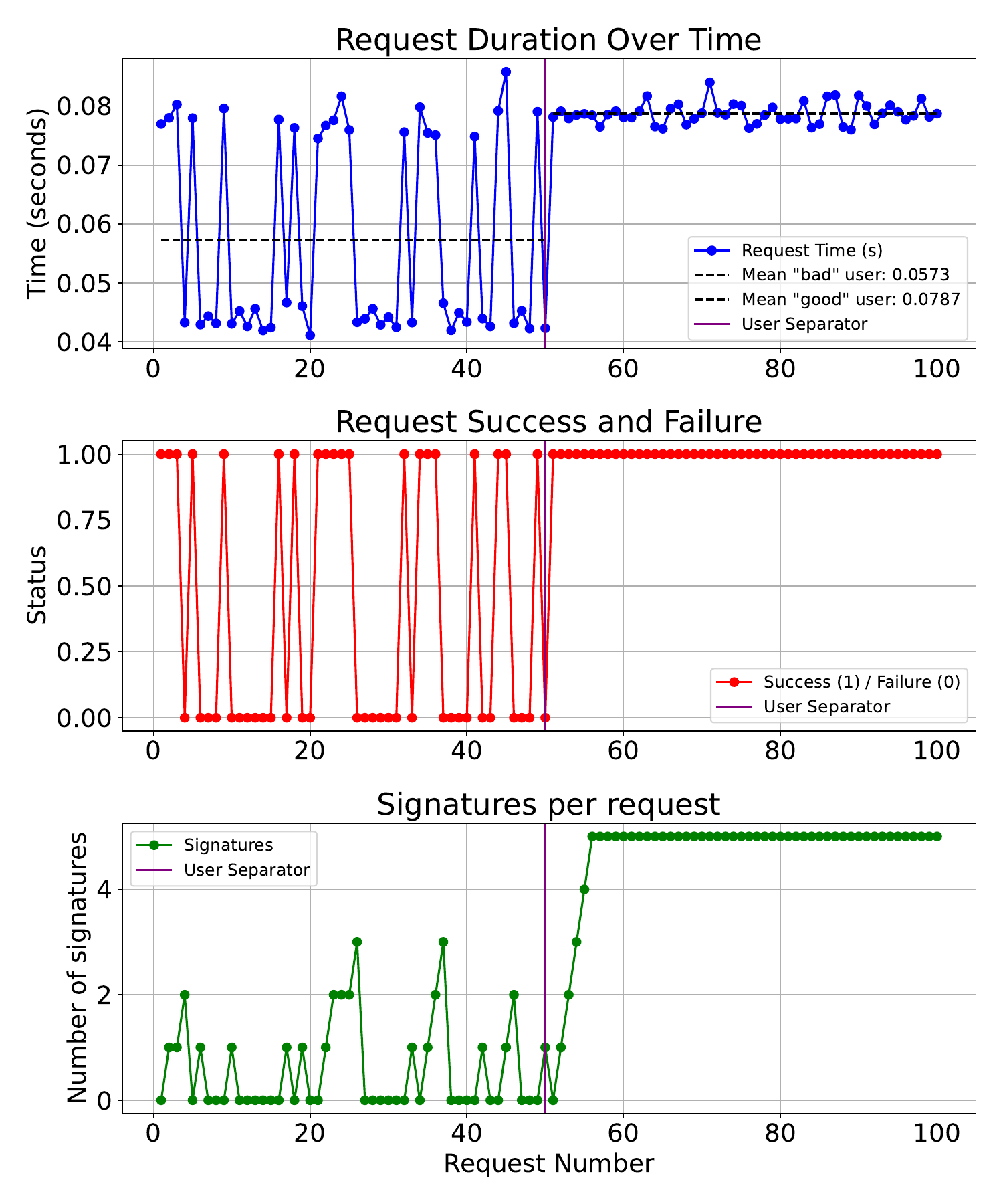}
  \centering
  \caption{First experiment}
  \label{fig:experiment}
\end{figure}

\begin{figure}
\begin{subfigure}{.5\textwidth}
  \centering
  \includegraphics[width=.8\linewidth]{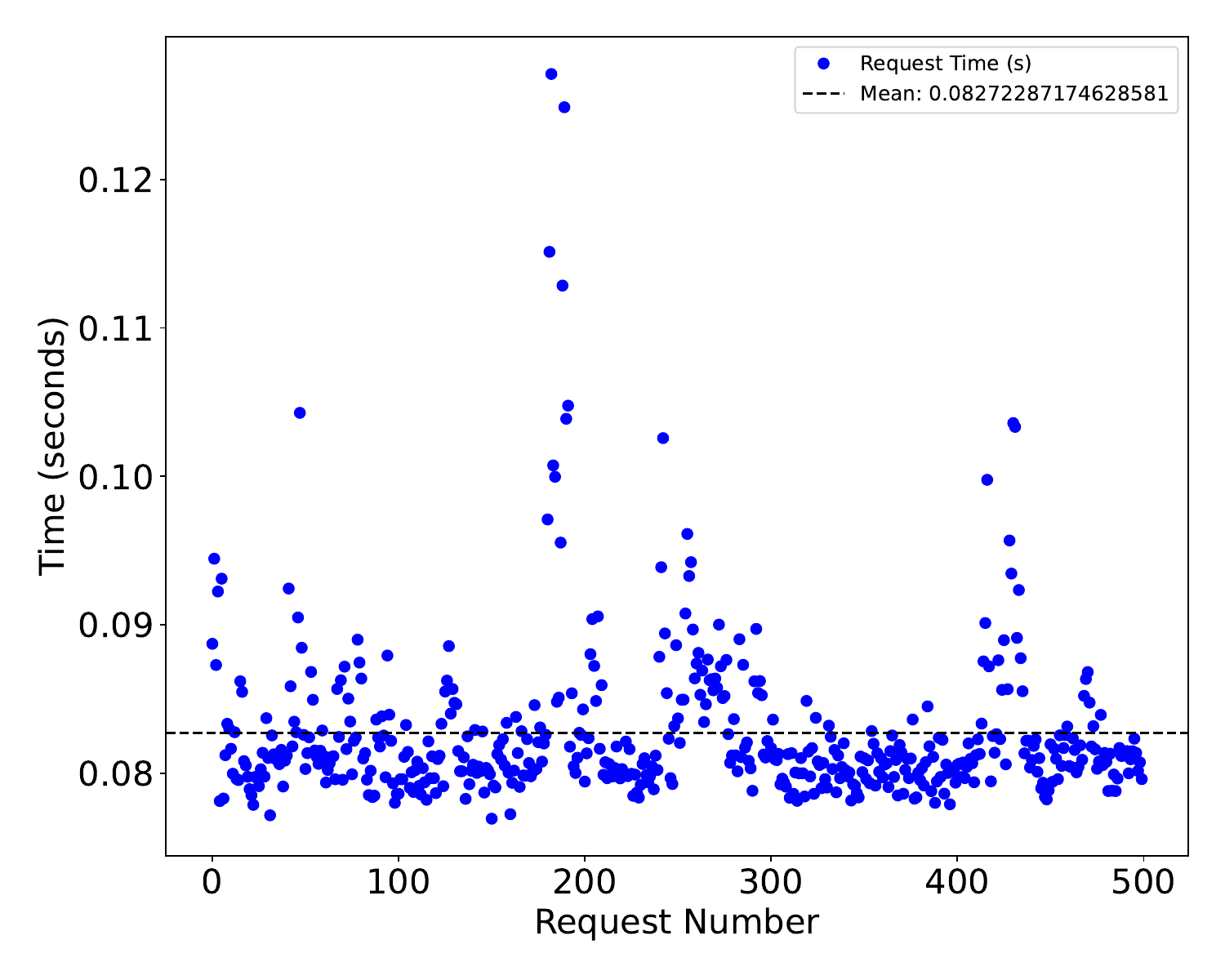}
  \caption{0 signatures user}
  \label{fig:sfig1}
\end{subfigure}
\\
\begin{subfigure}{.5\textwidth}
  \centering
  \includegraphics[width=.8\linewidth]{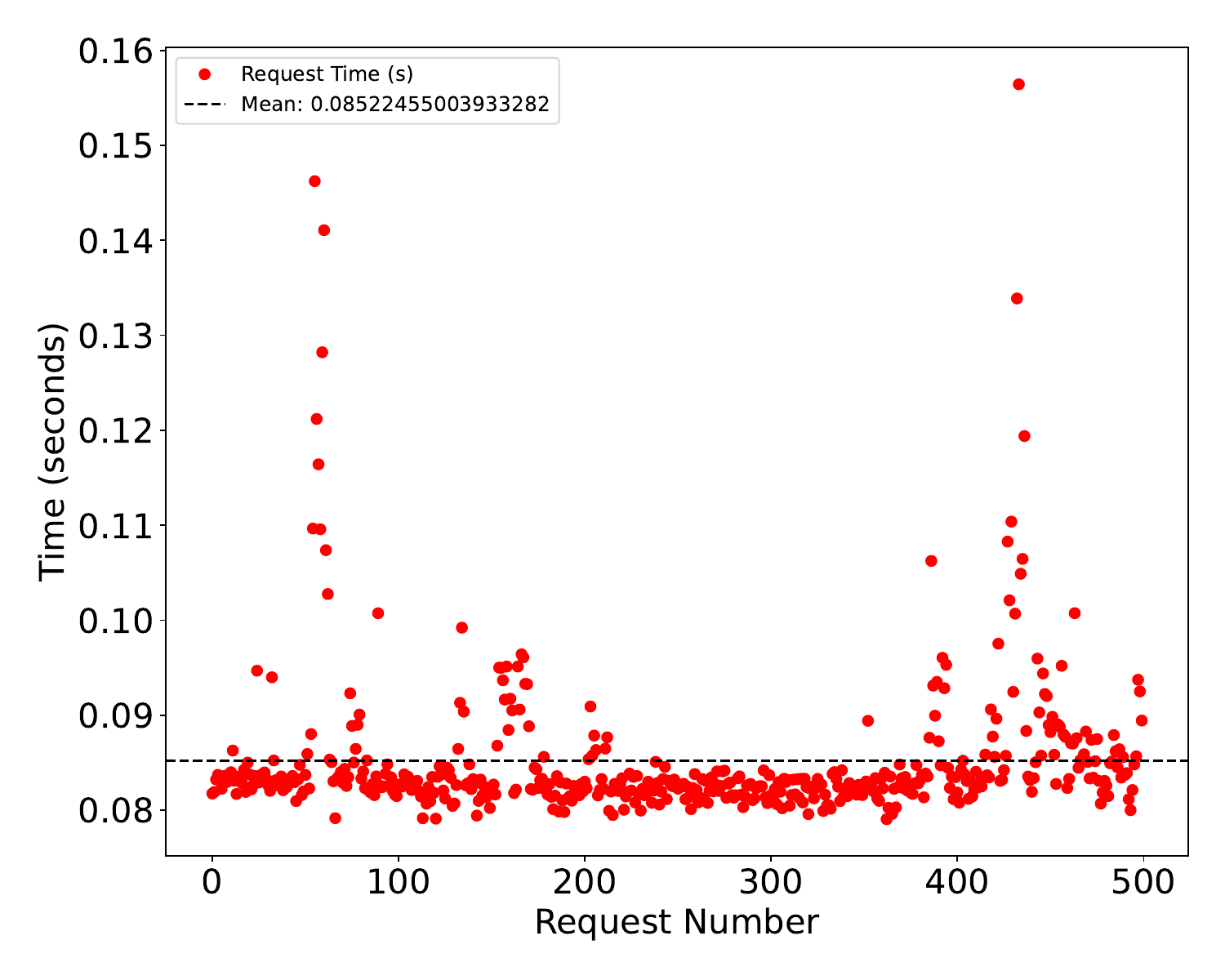}
  \caption{5 signatures user}
  \label{fig:sfig2}
\end{subfigure}
\caption{Signature latency}
\label{fig:latency_sign}
\end{figure}

While the usage of the score is not implemented in this POC, it can be stored on every server as a table of public keys with their scores, every server having a snapshot of the network at a certain point \ref{table:1}. ModSecurity has multiple score-based parameters(anomaly level, paranoia level) that can be set based on the request score to treat user based on their reputation.
\begin{table}[h]
\begin{center}
\begin{tabular}{ |c|c| } 
 \hline
 Public Key & Score \\ 
 \hline \hline
 PK1 & 5 \\ 
 \hline
 PK2 & 2 \\
 \hline
 PK3 & 3 \\
 \hline
 PK4 & 0 \\
 \hline
\end{tabular}
\end{center}
\caption{Table of public key scores.}
\label{table:1}
\end{table}

TrustZero uses the RSA cryptographic algorithm to create the public/private pair with a public key size of 256 bytes. The signature resulting from it inherits this dimension and creates a linear growth. To address a potential size issue, other signature algorithms such as Elliptic curve P-256(64 bytes public key) or EdDSA Ed25519(32 bytes public key) can be used. Both options are present in the Cryptography library in Python resulting in the same size of signature of 64 bytes. Adding the public key itself to the token dimension, the comparison, calculated up to 100 signatures, is presented in Figure \ref{fig:sizes}. As highlighted in future experiments, the signatures did not present a significant overhead in communication and can also be limited to a maximum number to resolve the potential drawback. All the following requests and investigations are produced using simple RSA signatures.   

For the experiments, 5 servers were deployed with docker, each container consuming as much CPU as it needed from the host. This setup was hosted on 2 different specifications: low resources (8 cores Intel(R) Core(TM) i7-8565U CPU @ 1.80GHz) and high resources(16 cores AMD Ryzen 7 6800H 4.7 GHz). All the experiments were successfully run on both specifications showing the low overhead of TrustZero even with a low-grade CPU. The subsequent results were used from the best-performing setup. One of the first experiments was to test whether any attacker that is tampering with the signatures is detected and their requests are denied. Two users were created sequentially to communicate with all the servers sending 50 requests and the results collected from all instances are presented in \ref{fig:experiment}. In the second half of the figure, a genuine user is building the maximum trust over time(5) and, even though having a higher average latency, does not encounter a high variability in times and better predictability. In contrast, once a user behaves abnormally, like in the first half of the plot, and sends corrupted signatures he loses his reputation and access to the resources and needs to build back the trust by interacting with the servers. In addition, every time a token is sent with modified signatures, the application firewall refuses immediately the call before it gets to the server, observation reflected in the low latency present in unsuccessful requests.

ModSecurity provides multiple automated defense mechanisms, starting from rate limiting to IP blacklisting integration. In a previous work regarding attack mitigations using WAFs\cite{keijer2019automated}, ModSecurity has been deployed to protect a server against a DDoS attack. Using his integrated IP blocking functionality based on a text file access, at the header level the overhead was measured at 509 microseconds. In TrustZero architecture, the trust token present in a header is parsed to an external script that signals ModSecurity the successful processing. This creates a higher amount of latency as there is no integrated functionality to process headers and the whole request is transferred to a python script. Based on the logging provided by ModSecurity in processing rules,
\begin{lstlisting}
[/login][4] Operator completed in 79255 usec.
\end{lstlisting}
we observed that the number of signatures in a trust token is not influencing the processing time of the rule as presented in Table \ref{table:2}.

\begin{table}[h]
\begin{center}
\begin{tabular}{ |c|c| } 
 \hline
 \makecell{Number of \\ signatures} & \makecell{Processing time \\ (in microseconds)} \\ 
 \hline \hline
 0 & 79255 \\ 
 \hline
 1 & 87352 \\
 \hline
 2 & 85477 \\
 \hline
 3 & 73568 \\
 \hline
 4 & 85917 \\
 \hline
 5 & 81200 \\
 \hline
\end{tabular}
\end{center}
\caption{Processing time of signatures}
\label{table:2}
\end{table}
To better understand the implication of signatures in response latency, measurements of time were taken from 2 specific cases: a user with all 5 server signatures and one with no reputation. 500 requests were created and resolved and the difference in computational time was around $0.025$ seconds, presenting a low latency of adding signatures in the request.

\begin{figure}[h]
  \includegraphics[width=.8\linewidth]{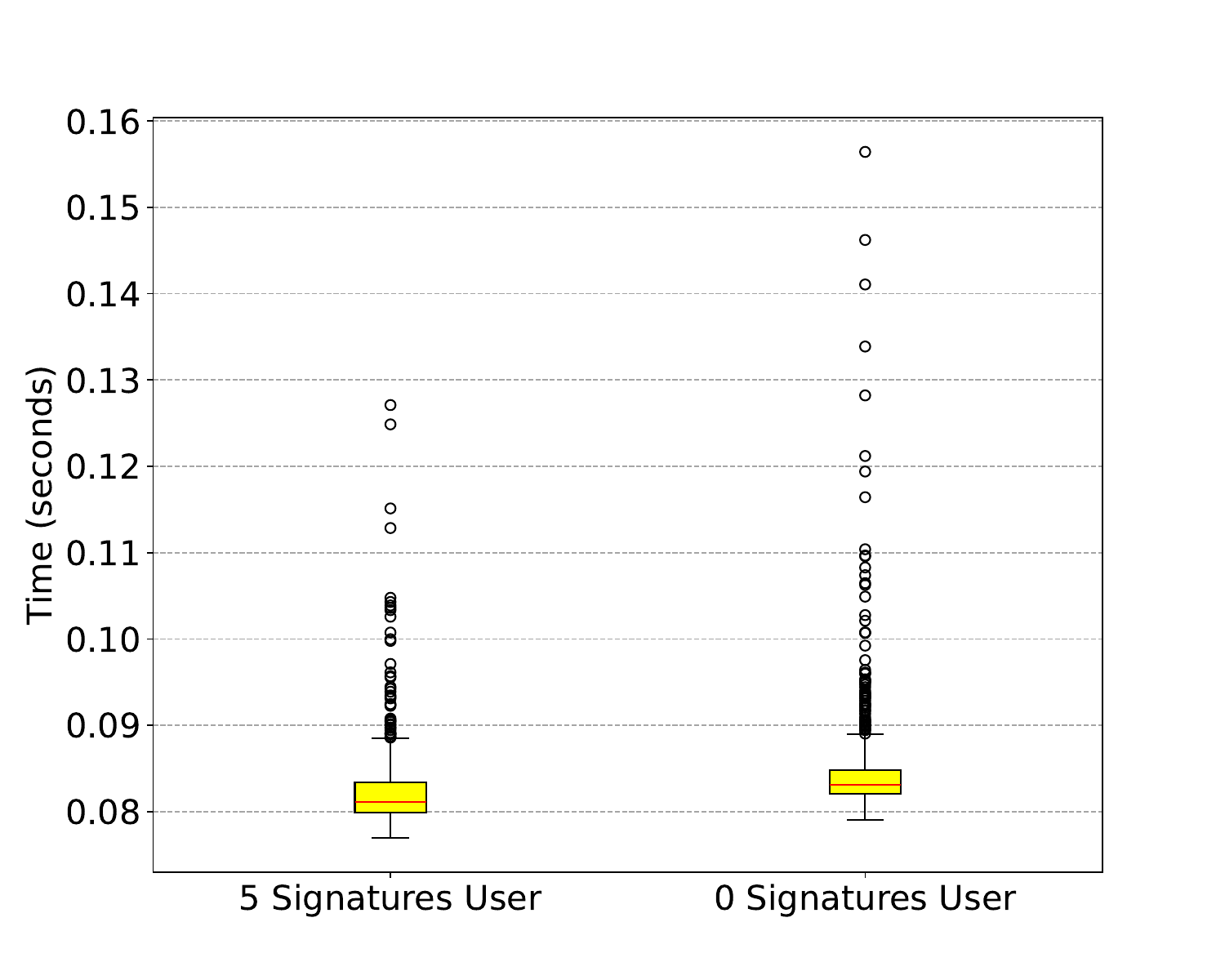}
  \centering
  \caption{Box plot signatures latency}
  \label{fig:latency_box}
\end{figure}

A box plot \ref{fig:latency_box} was created based on the latency set measured for better visibility of the data and its spreading. From the figure, both groups have similar distributions, but the user with more signatures shows a slightly higher median and fewer outliers compared to the plain one. In both cases, the number of values above the 95th percentile is small compared to the number of total requests reflecting a low variability over a large number of messages.

\begin{figure}[h]
  \includegraphics[width=.9\linewidth]{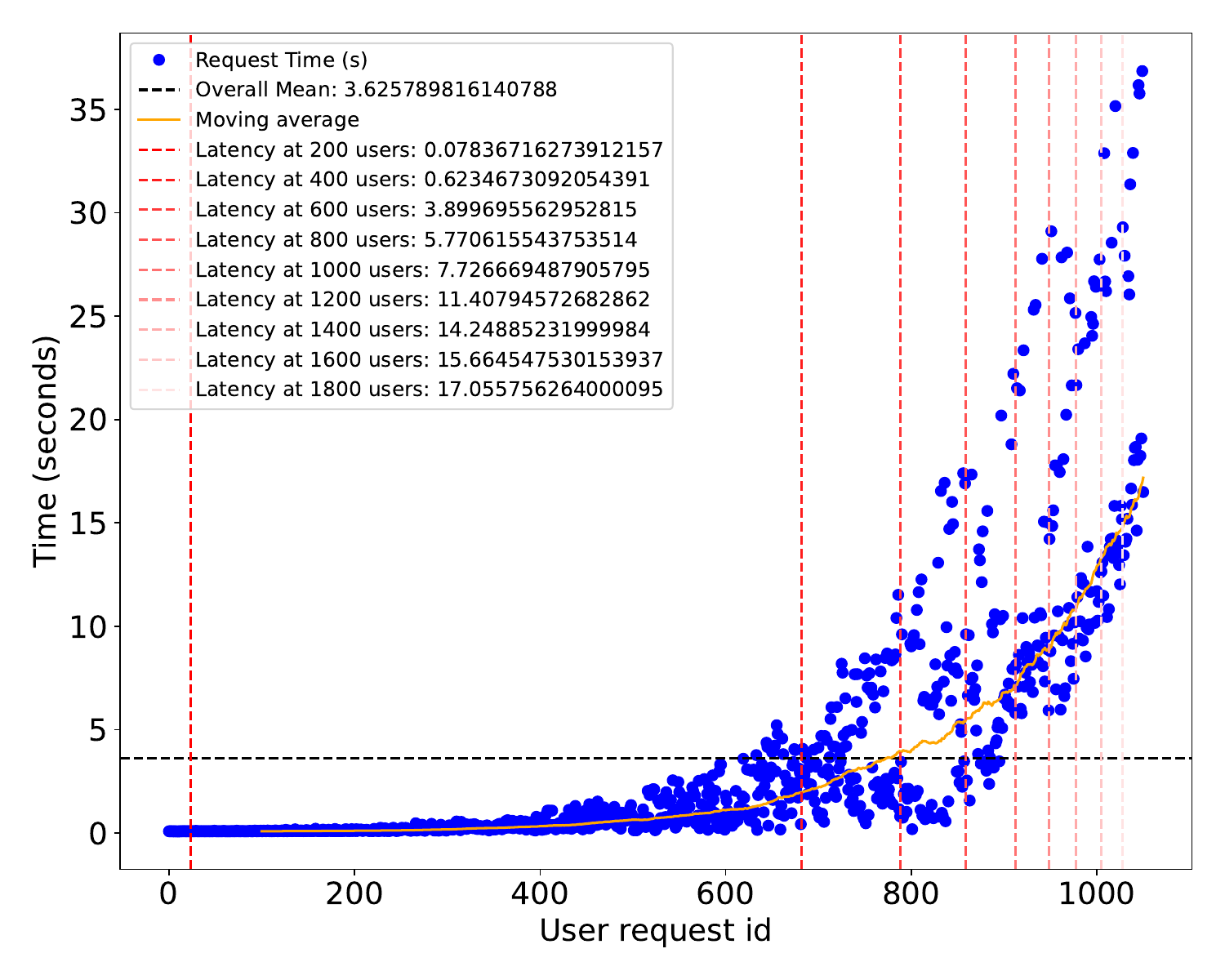}
  \centering
  \caption{Latency with up to 2000 users}
  \label{fig:latency_users}
\end{figure}

The largest experiment measured was deploying in the network up to 2000 users in different threads, adding every 2 seconds a new instance and measuring the latency of a legitimate user \ref{fig:latency_users}. This experiment can be considered as a "DDoS" attack launched over time trying to collapse the servers with repetitive and identical requests. As all instances are created in the same computer(including servers), the CPU and memory usage might influence the request time as more users flood the network. The test measurements were split after every 200 users were added and 3 phases can be identified:
\begin{enumerate}
    \item For the first 300 users the request times are relatively constant without big spikes of latency keeping the latency below the experiment mean
    \item From 300 to 1800, the latencies fluctuate significantly more and increase in mean time exponentially
    \item After 1800 users, the requests are resolved slowly (up to 35 seconds); the minimum latency is significantly higher than the other phases
\end{enumerate}
Moreover, for the first 200 users, the request times reported stay consistent or lower with the measurements  of previous experiments(of around $0.07$ seconds). A considerable mass of low latencies is still present at up to 600 users. For better visualization, a moving average trend line was created to fit the growth tendency, being updated with a sliding window of 100 requests.

TrustZero was developed with a focus on transparency and ease of integration. Recognizing the importance of seamless interoperability between organizations, TrustZero introduces a streamlined way to enhance your security without adding unnecessary complexity.

To achieve this, we developed a proof-of-concept Android application designed to integrate effortlessly with any existing app. This tool allows you to automatically include your public key in the headers of outgoing traffic, ensuring secure identification and communication.
TrustZero enables your application to manage received signatures effectively, including storing them securely and modifying them as needed to align with your security protocols and changes. The example of the integration app is presented in the annex \ref{FirstAppendix} with the code available in the same \href{https://github.com/AdiDumi/VPNApp}{repository}\cite{reposit}. Moreover, to test the actual integration of the keys in communication, the app was enhanced with an experiment where, after the containers(representing the servers) were started, a user could send requests and store the signatures received from different servers. This represents the final end-to-end open-source experiment towards passport-level trust including identity(key) generation and token exchange.

\begin{figure}[h]
  \includegraphics[width=\linewidth]{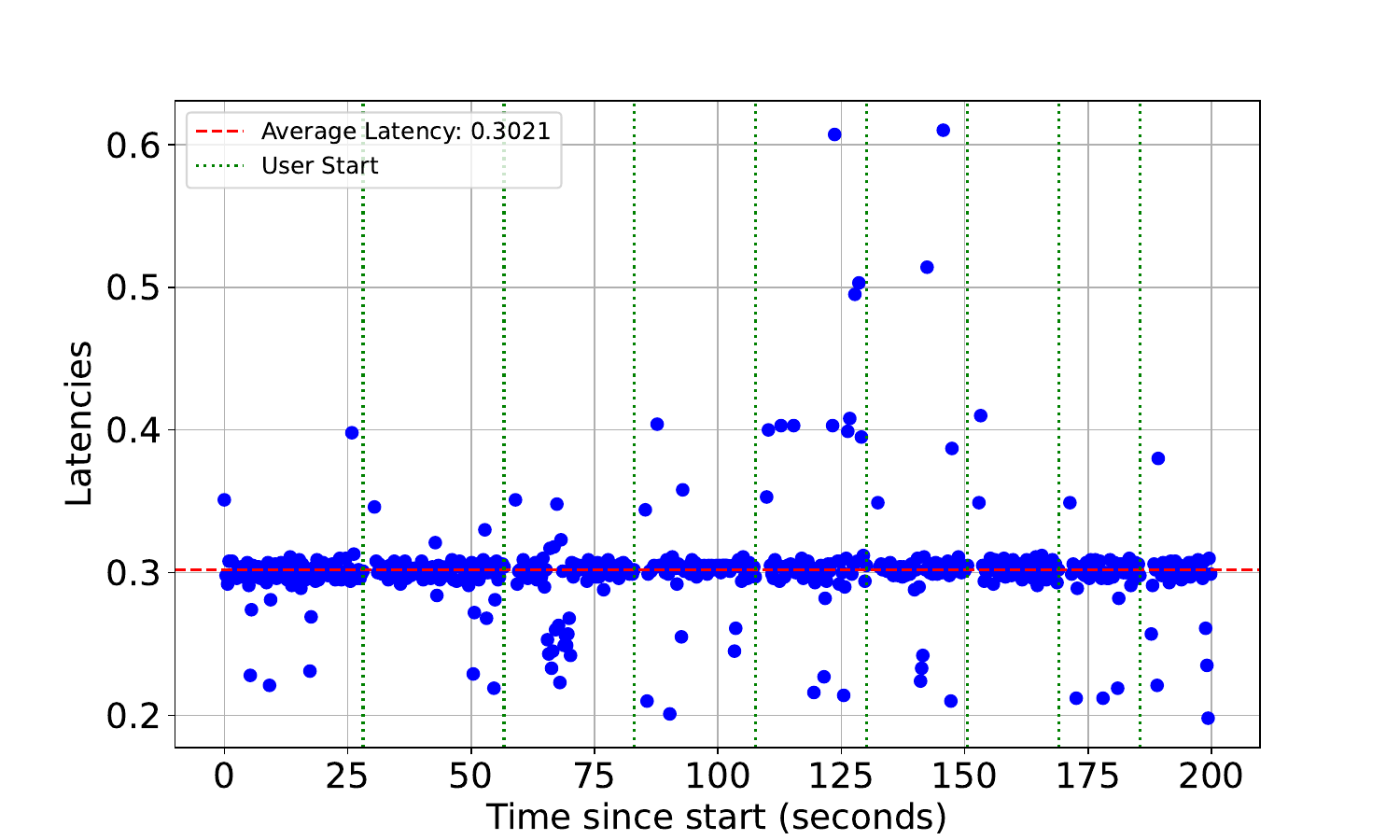}
  \centering
  \caption{Latency of 9 Android users}
  \label{fig:phone}
\end{figure}

\begin{figure}[h]
    \includegraphics[width=.24\textwidth]{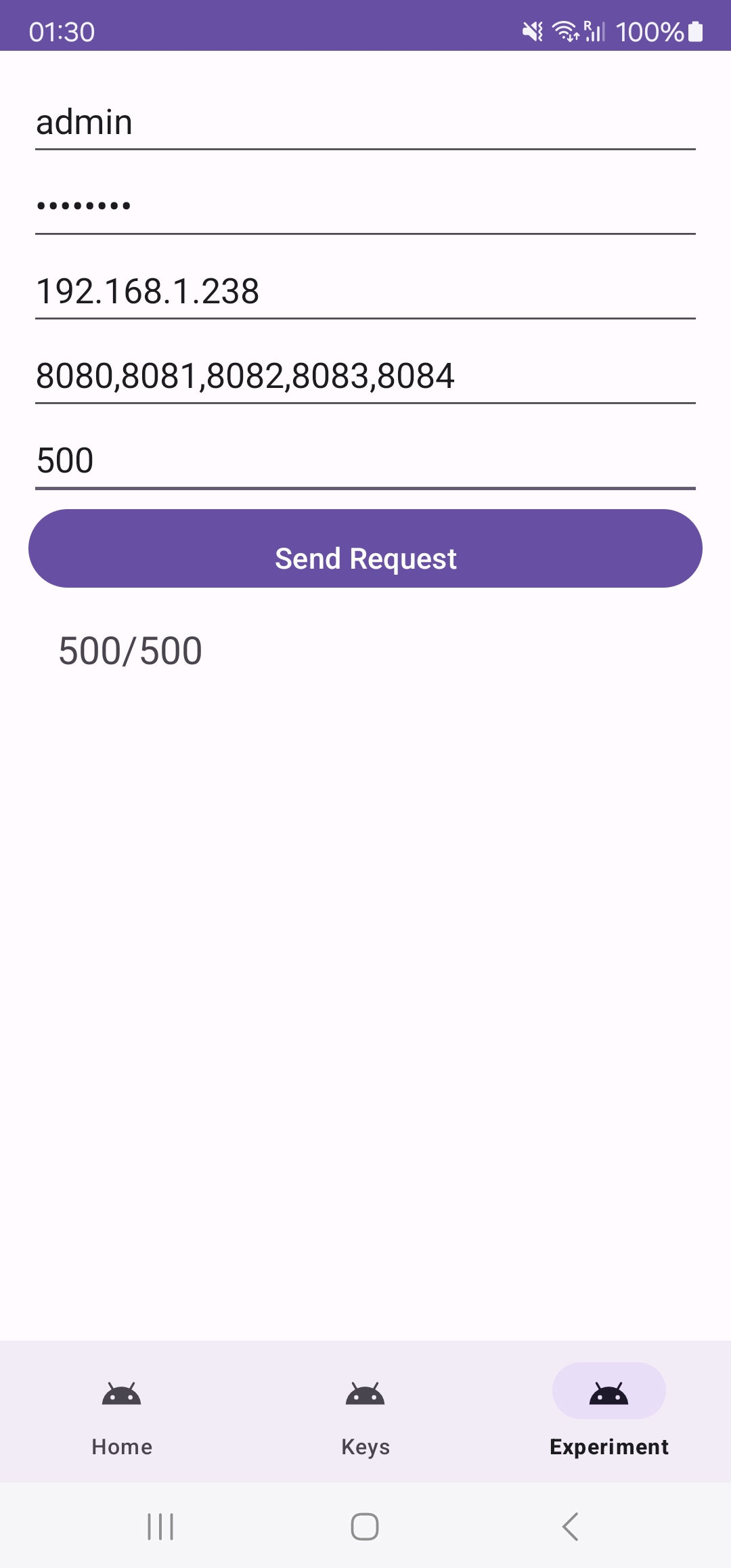}\hfill
    \includegraphics[width=.24\textwidth]{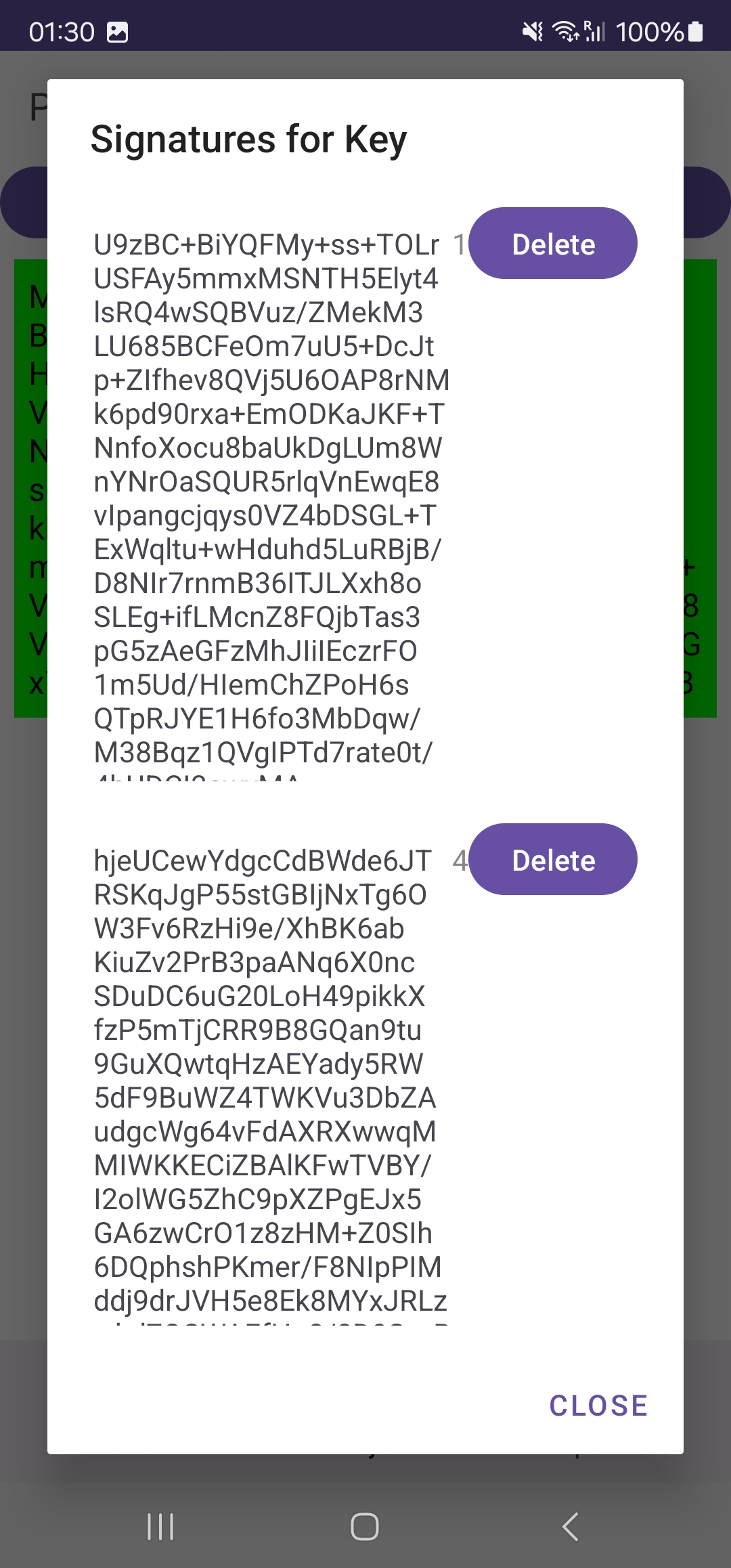}\hfill
    \caption{Android TrustZero experiment}\label{fig:exp}
\end{figure}

An initial performance analysis was measured with 9 Android phones starting to send requests to the servers over time. The results from Figure \ref{fig:phone} show an overall higher average at $0.3$ seconds but with a small interval distribution. This can be explained by the additional routing actions executed between the phone and the containerized servers running on the localhost machine. The UI of the experiment page alongside the resulting signatures received by 1 user are available in the annex at figure \ref{fig:exp}.

\section{Conclusion}

In an increasingly interconnected and volatile digital landscape, Zero Trust Architecture (ZTA) represents a vital paradigm shift in cybersecurity. This thesis introduced TrustZero, a scalable zero-trust security framework designed to address the limitations of traditional models. By leveraging a universal trust token and integrating robust cryptographic principles, TrustZero enhances trust portability, enables secure inter-organizational communication, and provides a resilient, mathematically grounded framework.

Our research underscores the transformative potential of combining zero-trust principles with lightweight cryptographic techniques to balance security and usability. Through rigorous testing, we demonstrated the feasibility of a distributed trust model that adapts to real-world complexities, such as supply chain interdependencies and evolving cyber threats. Experiments validated the efficiency of the trust scoring mechanism, showing minimal latency impacts and practical applicability even under simulated denial-of-service conditions.

TrustZero addresses the gap of implicit trust, as envisioned by ZTA, by focusing on mitigating the cost of breaches through continuous authentication and dynamic trust scoring. We implemented a proof-of-concept where we measured the latency of user requests and the resilience of the server in cases of HTTP floodings. Combining the zero-trust principle
and cryptography in our design we create a strong identity and web-of-trust framework suitable to mitigate not only network attacks but also interferences in social activities.

Ultimately, TrustZero provides a foundational step toward an open, verifiable, and scalable zero-trust security ecosystem. By emphasizing transparency and reproducibility, this framework not only advances academic and industrial scenarios but also sets a precedent for future developments in cybersecurity.

\bibliographystyle{plain}
\bibliography{ref}

\begin{thebibliography}{10}

\bibitem{europaCommissionOnline}
{C}ommission, online platforms and civil society increase monitoring during {R}omanian elections --- ec.europa.eu.
\newblock \url{https://ec.europa.eu/commission/presscorner/detail/en/ip_24_6243}.
\newblock [Accessed 10-12-2024].

\bibitem{europaCommissionOpens}
{C}ommission opens formal proceedings against {T}ik{T}ok on election risks under the {D}igital {S}ervices {A}ct --- ec.europa.eu.
\newblock \url{https://ec.europa.eu/commission/presscorner/detail/en/ip_24_6487}.
\newblock [Accessed 17-12-2024].

\bibitem{openzitiHardwareSecurity}
{H}ardware {S}ecurity {M}odules | {O}pen{Z}iti --- openziti.io.
\newblock \url{https://openziti.io/docs/guides/hsm/#enabling-a-ziti-endpoint-using-an-hsm}.
\newblock [Accessed 04-10-2024].

\bibitem{audiResponsibilitySupply}
{R}esponsibility in the supply chain | audi.com --- audi.com.
\newblock \url{https://www.audi.com/en/sustainability/people-society/responsibility-in-the-supply-chain.html}.
\newblock [Accessed 01-12-2024].

\bibitem{asmlResponsibleSupply}
{R}esponsible supply chain - {W}orking with our suppliers to become a sustainable leader in our industry --- asml.com.
\newblock \url{https://www.asml.com/en/company/sustainability/responsible-supply-chain}.
\newblock [Accessed 12-10-2024].

\bibitem{politicoRomaniasPresidential}
{R}omania’s presidential front-runner {G}eorgescu benefited from {R}ussia-style booster campaign, declassified docs say --- politico.eu.
\newblock \href{https://www.politico.eu/article/romanias-presidential-frontrunner-benefited-from-russia-style-booster-campaign-declassified-docs-say/}{https://www.politico.eu/article/romanias-presidential-frontrunner-benefited-from-russia-style-booster-campaign-declassified-docs-say/}.
\newblock [Accessed 16-12-2024].

\bibitem{reposit}
Trustzero.
\newblock \url{https://github.com/AdiDumi/TrustZero}.

\bibitem{openzitiWhatOpenZiti}
{W}hat is {O}pen{Z}iti? | {O}pen{Z}iti --- openziti.io.
\newblock \url{https://openziti.io/docs/learn/introduction/}.
\newblock [Accessed 04-10-2024].

\bibitem{eubook}
{ European Systemic Risk Board}.
\newblock {\em Advancing macroprudential tools for cyber resilience – Operational policy tools}.
\newblock European Systemic Risk Board, 2024.

\bibitem{tiberNL}
AFM.
\newblock Tiber-nl programme, 2016.

\bibitem{akbar2018sql}
Memen Akbar, Muhammad Arif~Fadhly Ridha, et~al.
\newblock Sql injection and cross site scripting prevention using owasp modsecurity web application firewall.
\newblock {\em JOIV: International Journal on Informatics Visualization}, 2(4):286--292, 2018.

\bibitem{10646352}
Faris Alsulami, Akshay~R. Kulkarni, Noor~Ahmad Hazari, and Mohammed~Y. Niamat.
\newblock Zebra: Zero trust architecture employing blockchain technology and ropuf for ami security.
\newblock {\em IEEE Access}, 12:119868--119883, 2024.

\bibitem{birch2017fraud}
Sarah Birch and Fatma ElSafoury.
\newblock Fraud, plot, or collective delusion? social media and perceptions of electoral misconduct in the 2014 scottish independence referendum.
\newblock {\em Election Law Journal}, 16(4):470--484, 2017.

\bibitem{chen2020security}
Baozhan Chen, Siyuan Qiao, Jie Zhao, Dongqing Liu, Xiaobing Shi, Minzhao Lyu, Haotian Chen, Huimin Lu, and Yunkai Zhai.
\newblock A security awareness and protection system for 5g smart healthcare based on zero-trust architecture.
\newblock {\em IEEE internet of things journal}, 8(13):10248--10263, 2020.

\bibitem{chigada2021cyberattacks}
Joel Chigada and Rujeko Madzinga.
\newblock Cyberattacks and threats during covid-19: A systematic literature review.
\newblock {\em South African Journal of Information Management}, 23(1):1--11, 2021.

\bibitem{dinculeanua2019vulnerabilities}
Dan Dinculean{\u{a}} and Xiaochun Cheng.
\newblock Vulnerabilities and limitations of mqtt protocol used between iot devices.
\newblock {\em Applied Sciences}, 9(5):848, 2019.

\bibitem{escobedo2017beyondcorp}
Victor Escobedo, Betsy Beyer, Max Saltonstall, and Filip Zyzniewski.
\newblock Beyondcorp: The user experience.
\newblock {\em Login}, 42(3):38--43, 2017.

\bibitem{tiberEU}
{European Central Bank}.
\newblock Tiber-eu framework, 2018.

\bibitem{feghhi2016web}
Saman Feghhi and Douglas~J Leith.
\newblock A web traffic analysis attack using only timing information.
\newblock {\em IEEE Transactions on Information Forensics and Security}, 11(8):1747--1759, 2016.

\bibitem{ferretti2021survivable}
Luca Ferretti, Federico Magnanini, Mauro Andreolini, and Michele Colajanni.
\newblock Survivable zero trust for cloud computing environments.
\newblock {\em Computers \& Security}, 110:102419, 2021.

\bibitem{fidler2014anarchy}
Mailyn Fidler.
\newblock Anarchy or regulation: Controlling the global trade in zero-day vulnerabilities.
\newblock {\em PhD diss., Freeman Spogli Institute for International Studies, Stanford University}, 2014.

\bibitem{frigieri2015m2m}
Edielson~P Frigieri, Daniel Mazzer, and LFCG Parreira.
\newblock M2m protocols for constrained environments in the context of iot: A comparison of approaches.
\newblock In {\em International Telecommunications Symposium}, page~5. sn, 2015.

\bibitem{gligor2022zero}
Virgil~D Gligor.
\newblock Zero trust in zero trust.
\newblock Technical report, CMU CyLab Technical Report 22--002 December 17, 2022.

\bibitem{gonccalves2023beyondcorp}
Guilherme Gon{\c{c}}alves, Kyle O'Malley, Max Saltonstall, et~al.
\newblock Beyondcorp and the long tail of zero trust.
\newblock 2023.

\bibitem{guo2023intelligent}
Xian Guo, Hongbo Xian, Tao Feng, Yongbo Jiang, Di~Zhang, and Junli Fang.
\newblock An intelligent zero trust secure framework for software defined networking.
\newblock {\em PeerJ Computer Science}, 9:e1674, 2023.

\bibitem{he2022survey}
Yuanhang He, Daochao Huang, Lei Chen, Yi~Ni, and Xiangjie Ma.
\newblock A survey on zero trust architecture: Challenges and future trends.
\newblock {\em Wireless Communications and Mobile Computing}, 2022(1):6476274, 2022.

\bibitem{hiesgen2022race}
Raphael Hiesgen, Marcin Nawrocki, Thomas~C Schmidt, and Matthias W{\"a}hlisch.
\newblock The race to the vulnerable: Measuring the log4j shell incident.
\newblock {\em arXiv preprint arXiv:2205.02544}, 2022.

\bibitem{8711673}
Trapti Jain and Nakul Jain.
\newblock Framework for web application vulnerability discovery and mitigation by customizing rules through modsecurity.
\newblock In {\em 2019 6th International Conference on Signal Processing and Integrated Networks (SPIN)}, pages 643--648, 2019.

\bibitem{janoky2018analysis}
L{\'a}szl{\'o}~Viktor J{\'a}noky, J{\'a}nos Levendovszky, and P{\'e}ter Ekler.
\newblock An analysis on the revoking mechanisms for json web tokens.
\newblock {\em International Journal of Distributed Sensor Networks}, 14(9):1550147718801535, 2018.

\bibitem{kaster2023privatized}
Sean~D Kaster and Prescott~C Ensign.
\newblock Privatized espionage: Nso group technologies and its pegasus spyware.
\newblock {\em Thunderbird International Business Review}, 65(3):355--364, 2023.

\bibitem{keijer2019automated}
Julik~S Keijer.
\newblock Automated ddos mitigation based on known attacks using a web application firewall.
\newblock {B.S.} thesis, University of Twente, 2019.

\bibitem{kindervag2010build}
John Kindervag et~al.
\newblock Build security into your network’s dna: The zero trust network architecture.
\newblock {\em Forrester Research Inc}, 27:1--16, 2010.

\bibitem{lakhno2022experimental}
V~Lakhno, A~Blozva, D~Kasatkin, V~Chubaievskyi, Y~Shestak, D~Tyshchenko, and R~Brzhanov.
\newblock Experimental studies of the features of using waf to protect internal services in the zero trust structure.
\newblock {\em J Theor Appl Inf Technol}, 100(3):705--721, 2022.

\bibitem{lampson2009privacy}
Butler Lampson.
\newblock Privacy and security usable security: how to get it.
\newblock {\em Communications of the ACM}, 52(11):25--27, 2009.

\bibitem{liu2011trustguard}
Haiqin Liu, Yan Sun, Victor~C Valgenti, and Min~Sik Kim.
\newblock Trustguard: A flow-level reputation-based ddos defense system.
\newblock In {\em 2011 IEEE Consumer Communications and Networking Conference (CCNC)}, pages 287--291. IEEE, 2011.

\bibitem{meghdouri2018analysis}
Fares Meghdouri, Tanja Zseby, and F{\'e}lix Iglesias.
\newblock Analysis of lightweight feature vectors for attack detection in network traffic.
\newblock {\em Applied Sciences}, 8(11):2196, 2018.

\bibitem{9104214}
Saima Mehraj and M.~Tariq Banday.
\newblock Establishing a zero trust strategy in cloud computing environment.
\newblock In {\em 2020 International Conference on Computer Communication and Informatics (ICCCI)}, pages 1--6, 2020.

\bibitem{naseer2024crowdstrike}
Iqra Naseer.
\newblock The crowdstrike incident: Analysis and unveiling the intricacies of modern cybersecurity breaches.
\newblock {\em World Journal of Advanced Engineering Technology and Sciences}, 10, 2024.

\bibitem{OCONNOR2013125}
TJ~O'Connor.
\newblock Chapter 4 - network traffic analysis with python.
\newblock In TJ~O'Connor, editor, {\em Violent Python}, pages 125--169. Syngress, 2013.

\bibitem{eudi}
GitHub~Organization of~the European Digital Identity~project.
\newblock European digital identity.
\newblock \url{https://github.com/eu-digital-identity-wallet/eudi-app-android-wallet-ui}, 2024.

\bibitem{ogundipe2024shaky}
Olugbenro Ogundipe and Tejiri Aweto.
\newblock The shaky foundation of global technology: A case study of the 2024 crowdstrike outage.
\newblock 2024.

\bibitem{peck2017migrating}
Jeff Peck, Betsy Beyer, Colin Beske, and Max Saltonstall.
\newblock Migrating to beyondcorp: maintaining productivity while improving security.
\newblock {\em Login}, 42(2):1--7, 2017.

\bibitem{petratos2014cybersecurity}
Pythagoras Petratos.
\newblock Cybersecurity in europe: Cooperation and investment.
\newblock {\em Cyber-Development, Cyber-Democracy and Cyber-Defense: Challenges, Opportunities and Implications for Theory, Policy and Practice}, pages 279--301, 2014.

\bibitem{rammal2024communication}
Ahmad Rammal, Kaja Gruntkowska, Nikita Fedin, Eduard Gorbunov, and Peter Richt{\'a}rik.
\newblock Communication compression for byzantine robust learning: New efficient algorithms and improved rates.
\newblock In {\em International Conference on Artificial Intelligence and Statistics}, pages 1207--1215. PMLR, 2024.

\bibitem{ristic2010modsecurity}
Ivan Ristic.
\newblock {\em Modsecurity handbook}.
\newblock Feisty Duck, 2010.

\bibitem{rogin2007cyber}
Josh Rogin.
\newblock Cyber officials: Chinese hackers attack ‘anything and everything,’.
\newblock {\em FCW. com, February}, 13:97658--1, 2007.

\bibitem{8473444}
Mayra Samaniego and Ralph Deters.
\newblock Zero-trust hierarchical management in iot.
\newblock In {\em 2018 IEEE International Congress on Internet of Things (ICIOT)}, pages 88--95, 2018.

\bibitem{santonisecurity}
Sebastian Santoni.
\newblock The security council.

\bibitem{6062066}
Fred~B. Schneider.
\newblock Beyond hacking: an sos!
\newblock In {\em 2010 ACM/IEEE 32nd International Conference on Software Engineering}, volume~1, pages 2--2, 2010.

\bibitem{9585170}
Malcolm Shore, Sherali Zeadally, and Astha Keshariya.
\newblock Zero trust: The what, how, why, and when.
\newblock {\em Computer}, 54(11):26--35, 2021.

\bibitem{singh2018impact}
Jatesh~Jagraj Singh, Hamman Samuel, and Pavol Zavarsky.
\newblock Impact of paranoia levels on the effectiveness of the modsecurity web application firewall.
\newblock In {\em 2018 1st International Conference on Data Intelligence and Security (ICDIS)}, pages 141--144. IEEE, 2018.

\bibitem{srivatsa2005trustguard}
Mudhakar Srivatsa, Li~Xiong, and Ling Liu.
\newblock Trustguard: countering vulnerabilities in reputation management for decentralized overlay networks.
\newblock In {\em Proceedings of the 14th international conference on World Wide Web}, pages 422--431, 2005.

\bibitem{stafford2020zero}
V~Stafford.
\newblock Zero trust architecture.
\newblock {\em NIST special publication}, 800:207, 2020.

\bibitem{9773102}
Naeem~Firdous Syed, Syed~W. Shah, Arash Shaghaghi, Adnan Anwar, Zubair Baig, and Robin Doss.
\newblock Zero trust architecture (zta): A comprehensive survey.
\newblock {\em IEEE Access}, 10:57143--57179, 2022.

\bibitem{teerakanok2021migrating}
Songpon Teerakanok, Tetsutaro Uehara, and Atsuo Inomata.
\newblock Migrating to zero trust architecture: Reviews and challenges.
\newblock {\em Security and Communication Networks}, 2021(1):9947347, 2021.

\bibitem{thapngam2011discriminating}
Theerasak Thapngam, Shui Yu, Wanlei Zhou, and Gleb Beliakov.
\newblock Discriminating ddos attack traffic from flash crowd through packet arrival patterns.
\newblock In {\em 2011 IEEE conference on computer communications workshops (INFOCOM WKSHPS)}, pages 952--957. IEEE, 2011.

\bibitem{vanickis2018access}
Romans Vanickis, Paul Jacob, Sohelia Dehghanzadeh, and Brian Lee.
\newblock Access control policy enforcement for zero-trust-networking.
\newblock In {\em 2018 29th Irish Signals and Systems Conference (ISSC)}, pages 1--6. IEEE, 2018.

\bibitem{ward2014beyondcorp}
Rory Ward and Betsy Beyer.
\newblock Beyondcorp: A new approach to enterprise security.
\newblock {\em ; login:: the magazine of USENIX \& SAGE}, 39(6):6--11, 2014.

\bibitem{yao2020dynamic}
Qigui Yao, Qi~Wang, Xiaojian Zhang, and Jiaxuan Fei.
\newblock Dynamic access control and authorization system based on zero-trust architecture.
\newblock In {\em Proceedings of the 2020 1st international conference on control, robotics and intelligent system}, pages 123--127, 2020.

\bibitem{zanasi2024flexible}
Claudio Zanasi, Silvio Russo, and Michele Colajanni.
\newblock Flexible zero trust architecture for the cybersecurity of industrial iot infrastructures.
\newblock {\em Ad Hoc Networks}, 156:103414, 2024.

\end{thebibliography}
\newpage
\newpage
\appendices
\section{First Appendix}
\label{FirstAppendix}
\begin{figure}[h]
    \includegraphics[width=.24\textwidth]{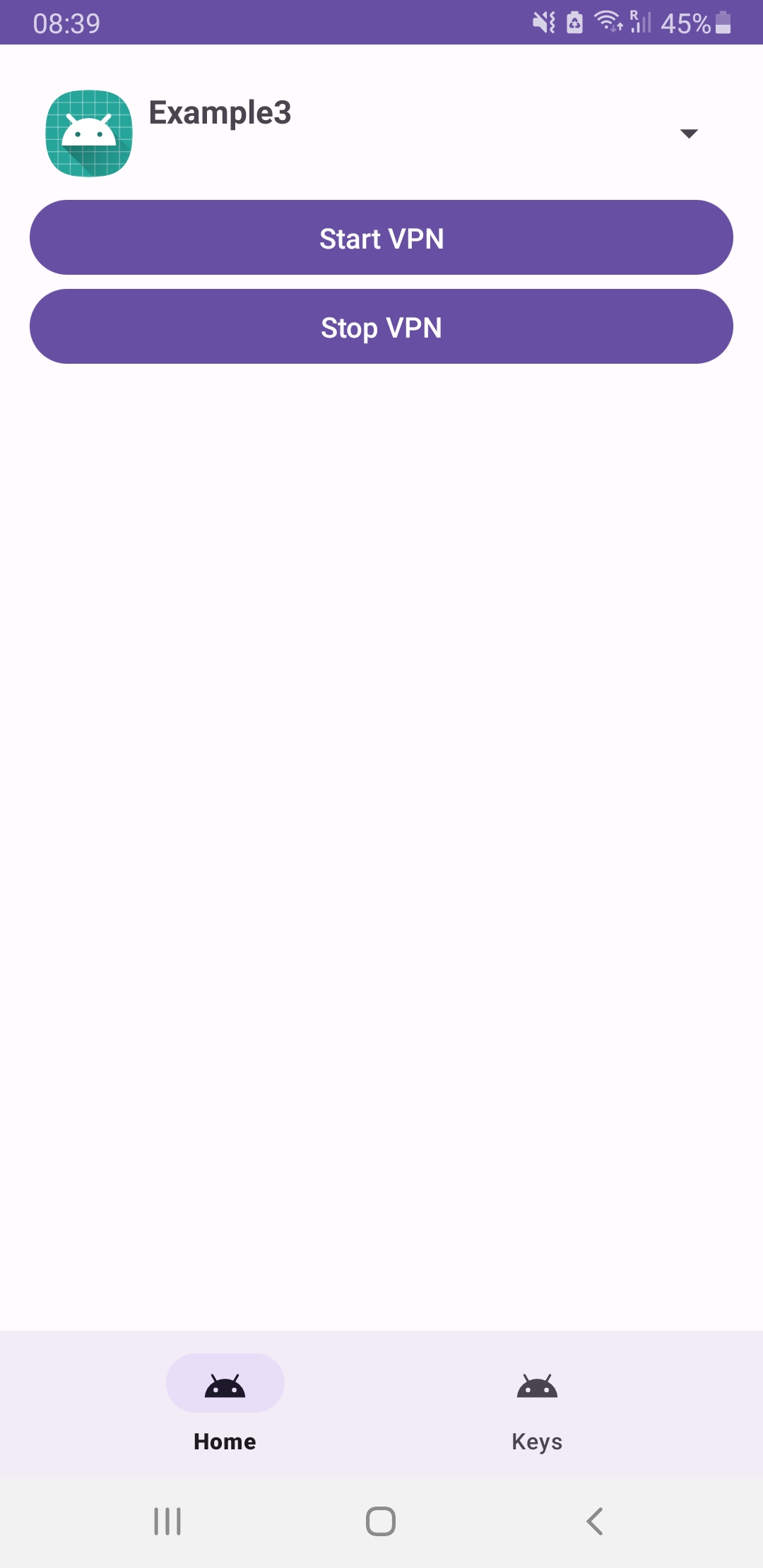}\hfill
    \includegraphics[width=.24\textwidth]{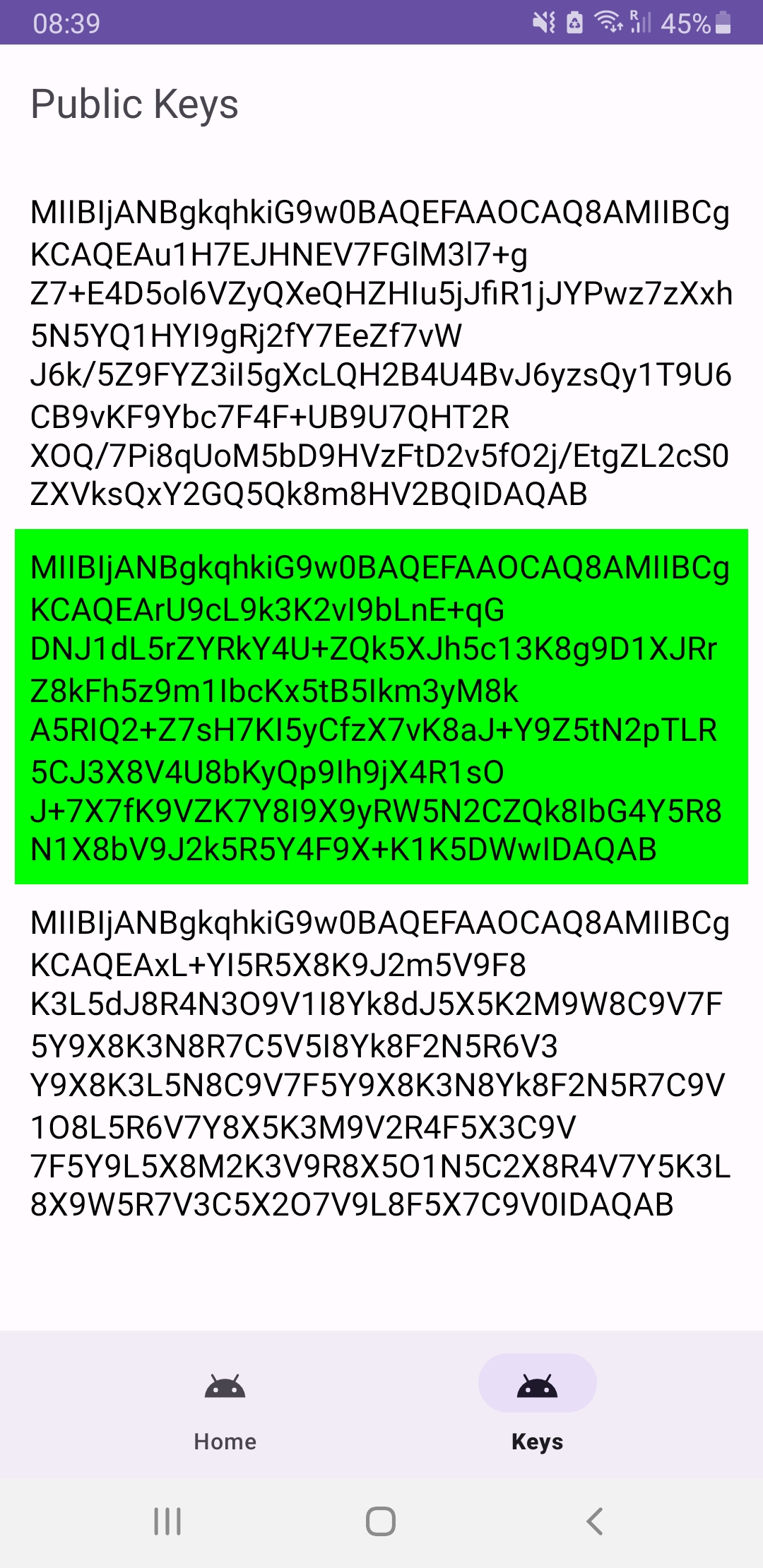}\hfill
    \caption{Android app for TrustZero integration}\label{fig:app}
\end{figure}

\end{document}